\documentclass[a4paper,11pt]{article}
\usepackage{jcappub} 
\usepackage{lineno}

\usepackage{comment}
\usepackage{subcaption}
\usepackage{bm}

\newcommand{\de}{\text{d}}
\newcommand{\Msun}{\text{M}_{\odot}}
\newcommand{\kmsec}{\text{km}/\text{s}}

\newcommand{\kpc}{\text{kpc}}
\newcommand{\pc}{\text{pc}} 

\newcommand{\Myr}{\text{Myr}}
\newcommand{\Gyr}{\text{Gyr}}
\newcommand{\eV}{\text{eV}}
\newcommand{\Msunppcc}{\Msun\pc^{-3}}

\newcommand{\masyr}{\text{mas}~\text{yr}^{-1}}

\title{\boldmath Fuzzy dark matter dynamics in tidally perturbed dwarf spheroidal galaxy satellites}

\author[a,b]{Axel Widmark,}
\author[c]{Tomer D. Yavetz,}
\author[d]{and Xinyu Li}
\affiliation[a]{Stockholm University and The Oskar Klein Centre for Cosmoparticle Physics,  Alba Nova, 10691 Stockholm, Sweden}
\affiliation[b]{Dark Cosmology Centre, Niels Bohr Institute, University of Copenhagen, Jagtvej 128, 2200 Copenhagen N, Denmark}
\affiliation[c]{School of Natural Sciences, Institute for Advanced Study, Princeton, NJ 08540, USA}
\affiliation[d]{Canadian Institute for Theoretical Astrophysics, 60 St George St, Toronto, ON M5R 2M8}

\emailAdd{axel.widmark@fysik.su.se}
\abstract{
Fuzzy dark matter (FDM) has dynamical properties that differ significantly from cold dark matter (CDM).
These dynamical differences are strongly manifested on the spatial scale of dwarf spheroidal galaxies (dSphs), which roughly corresponds to the de Broglie wavelength of a canonical mass FDM particle.
We study simulations of a dSph satellite which is tidally perturbed by its host galaxy, in order to identify dynamical signatures that are unique to FDM, and to quantify the imprints of such perturbations on an observable stellar tracer population.
We find that a perturbed FDM soliton develops a long-standing breathing mode, whereas for CDM such a breathing mode quickly phase-mixes and disappears. We also demonstrate that such signatures become imprinted on the dynamics of a stellar tracer population, making them observable with sufficiently precise astrometric measurements.
}

\begin{document}
\maketitle
\flushbottom

\section{Introduction}
\label{sec:intro}

Although the gravitational effects of dark matter (DM) have been extensively documented over the past decades, its true nature remains an enigma. Many ongoing efforts aimed at direct or indirect detection of DM have thus far turned up no evidence of the particles they seek to detect \cite{Liu2017,Schumann2019}. In the absence of DM particle detection, a promising method for differentiating between different DM candidates involves a detailed evaluation of their gravitational effects on the dynamics of baryonic tracer populations.

One DM candidate that has recently garnered increased attention is fuzzy dark matter (FDM), consisting of ultralight bosons with mass ranging between $\sim 10^{-23}$ eV and $\sim 10^{-20}$ eV \cite{Hui2017,Hui2021}. Owing to the ultralight particle mass, the main consequence of FDM is its quantum mechanical wave-like behaviour on astrophysical scales of up to several kiloparsecs. FDM has a larger Jeans length, which can prevent the formation of satellite halos and suppress the small scale matter power spectrum \cite{Hu2000}. This can lead to significant departures from the expected small-scale characteristics of standard cold dark matter (CDM), while retaining CDM’s successful predictions for large scale structure. The centre of an FDM halo is dominated by the ground state solution of the Schr\"odinger-Poisson equation, often referred to as a ``soliton,'' providing a natural explanation for the inner regions of galaxies exhibiting cored instead of cuspy density profiles \cite{Schive2014a}.

Recent studies have revealed that FDM exhibits other novel small scale features absent in the CDM picture, including the formation of vortices \cite{Hui2021b}, and oscillation and random walk of the central soliton \cite{Schwabe2016,Schive2020}. These phenomena originate from the interference between different eigenmodes of the wave function \cite{Li2021} and lead to density and gravity fluctuations of order unity. Cosmological simulations have demonstrated that the small scale structure caused by this interference pattern may increase the matter power spectrum below the de Broglie scale \cite{May2021}, casting doubt on previous constraints on FDM from the Lyman-$\alpha$ forest, derived from CDM simulations using a modified initial power spectrum with suppressed small scale power \cite{Irsic2017,Armengaud2017}. Meanwhile, the small scale dynamics open a new window to constrain FDM. Proposals include strong gravitation lensing \cite{Chan2020,Laroche2022}, tidal streams \cite{Dalal2021}, and heating of dwarf galaxies \cite{Marsh2019,Chiang2021,Dalal2022,DuttaChowdhury2023}.

In light of this uncertainty, a dynamical mechanism with observational consequences for differentiating between CDM and FDM in the present-day Universe would be of great value. The behaviour of DM in dwarf spheroidal galaxies (dSphs) and its effect on observable tracers are of particular interest, because dSphs are among the most DM dominated systems that we know of. They are also, in many ways, fairly simple systems. As such, they are potent laboratories for DM, for example in terms of indirect detection \cite{Conrad2014,Gaskins2016}, as well as the gravitational information that is available via their stellar dynamics.

For the lower end of the FDM mass range ($m_a \sim 10^{-22}$ eV), dSphs would likely be dominated by the central soliton. In an equilibrium scenario, an FDM soliton at rest is indistinguishable from the same mass distribution of steady state CDM. However, if we consider a time-varying scenario where the DM distribution has been disturbed by an external gravitational kick, then the dynamical response of FDM and CDM can be differentiated. Although often regarded as a liability, analysing non-equilibrium dynamics has been shown to be highly informative in other contexts, also when making dynamical mass measurements (e.g. weighing the Milky Way disk using the \emph{Gaia} phase-space spiral \cite{SpiralI,SpiralII,SpiralIII,SpiralIV}; constraining the phase-space distribution of the Milky Way DM halo from its response to being perturbed by the Large Magellanic Cloud \citep{Rozier2022}).

We focus on the internal non-equilibrium dynamics of a tidally perturbed dSph (rather than, for example, their stream-forming ejecta). This particular focus is timely, due to the current and near-future wealth of stellar dynamics data, for example coming from the astrometric \emph{Gaia} survey \cite{GaiaDR3}, radial velocity surveys (e.g. APOGEE \cite{Apogee2017}, SDSS-V's Milky Way Mapper \cite{SDSS-V_panoptic_spectro,sdss-v-dr18}), and integral field spectroscopy surveys (e.g. MaNGA \cite{Manga2016}, SDSS-V's Local Volume Mapper \cite{SDSS-V_panoptic_spectro,sdss-v-dr18}). Furthermore, the combination of different surveys can produce very precise proper motion measurements, especially in the case that observations epochs have a long time separation (e.g. combining \emph{Gaia} and the Hubble Space Telescope \cite{GaiaHSTcombined2022}). In short, observations of Local Group dSph satellites are becoming deeper and increasingly precise, constraining their orbits \cite{satorbits,Pace2022} and, in some cases, even their internal kinematics using proper motions \citep{Martinez-Garcia2021,Martinez-Garcia2023}. This new data has provided increasing evidence that some dSphs' internal kinematics are significantly perturbed by tidal forces (e.g. Hercules I \cite{Kupper2017,Mutlu-Pakdil2020,Gregory2020}, Crater II \cite{Fu2019,Borukhovetskaya2022}).

In this work, we run simulations of dSph satellites which are subject to a short tidal impulse during their pericentre passage, and compare the dynamical behaviour of FDM and CDM. We also simulate massless test-particles in the underlying DM potential and show that the time-varying DM dynamics also manifest themselves on an observable stellar tracer population.
We identify dynamical signatures that clearly differentiate FDM from CDM, mainly in the form of a persistent breathing mode. In general terms, this demonstrates that the non-equilibrium dynamics of dSphs serve as a potent probe of the dynamical properties of DM and its constituent particle.

This article is structured as follows. In section~\ref{sec:methods}, we describe our theoretical methods and the setup of our simulations. We then present our results in section~\ref{sec:results}. In sections~\ref{sec:discussion} and \ref{sec:conclusion}, we discuss and conclude.

\section{Theoretical methods and simulation setups}
\label{sec:methods}

We consider dSph satellites on eccentric orbits, which experience a short-lived tidal kick during their pericentre passage. In the FDM case, we assume that the envelope of excited states have already been stripped by tidal forces, leaving only a fully degenerate soliton. A similar scenario is found in ref.~\cite{Schive2020}; we refer especially to their figure 3, where the envelope of excited states has been tidally stripped while leaving the soliton core almost completely intact.

We study and compare the dynamical response of the FDM soliton with a corresponding CDM halo. In the FDM case, we also simulate massless test-particles, representing an observable stellar population. In the initial state of the simulation, the test-particles are in a steady state, with a number density that is proportional to the FDM matter density. The simulations are described in detail below.

\subsection{Coordinate system}\label{sec:coordinates}

The coordinate system used in this work is strictly in the rest-frame of the dSph satellite. We use a Cartesian grid of spatial coordinates, $\textbf{r} = (x,y,z)$, and their corresponding velocity coordinates, $(\dot{x},\dot{y},\dot{z}) \equiv (u,v,w)$.

The time coordinate ($t$), is zero at the point of pericentre passage, which is when the external tidal force is maximal. The direction of the tidal force rotates in the $(x,y)$-plane; at $t=0~\Myr$, its direction is along the $x$-axis.

In some of our visualisations, we show the phase-space distribution as it would be observed from the perspective of the host galaxy's centre, thus emulating what could realistically be observable for a Milky Way dSph satellite. For this purpose, we define a line-of-sight velocity,
\begin{equation}\label{eq:vlos}
    v_\text{l.o.s.} = \cos [a(t)] \, u + \sin [a(t)] \, v,
\end{equation}
and a spatial coordinate,
\begin{equation}\label{eq:xi}
    \xi = - \sin [a(t)] \, x + \cos [a(t)] \, y,
\end{equation}
which is parallel to the $(x,y)$-plane and perpendicular to the line-of-sight. Furthermore, we define a projected radius,
\begin{equation}\label{eq:rproj}
    r_\text{proj.} = \sqrt{\xi^2+z^2},
\end{equation}
which is the distance from the satellite's centre in the 2d plane that is perpendicular to the line-of-sight. These coordinates emulate what would be observable when the spatial coordinate along the line-of-sight is unavailable.

\subsection{External tidal force}\label{sec:tidal_force}

\begin{figure}
\centering
\includegraphics[width=.86\textwidth]{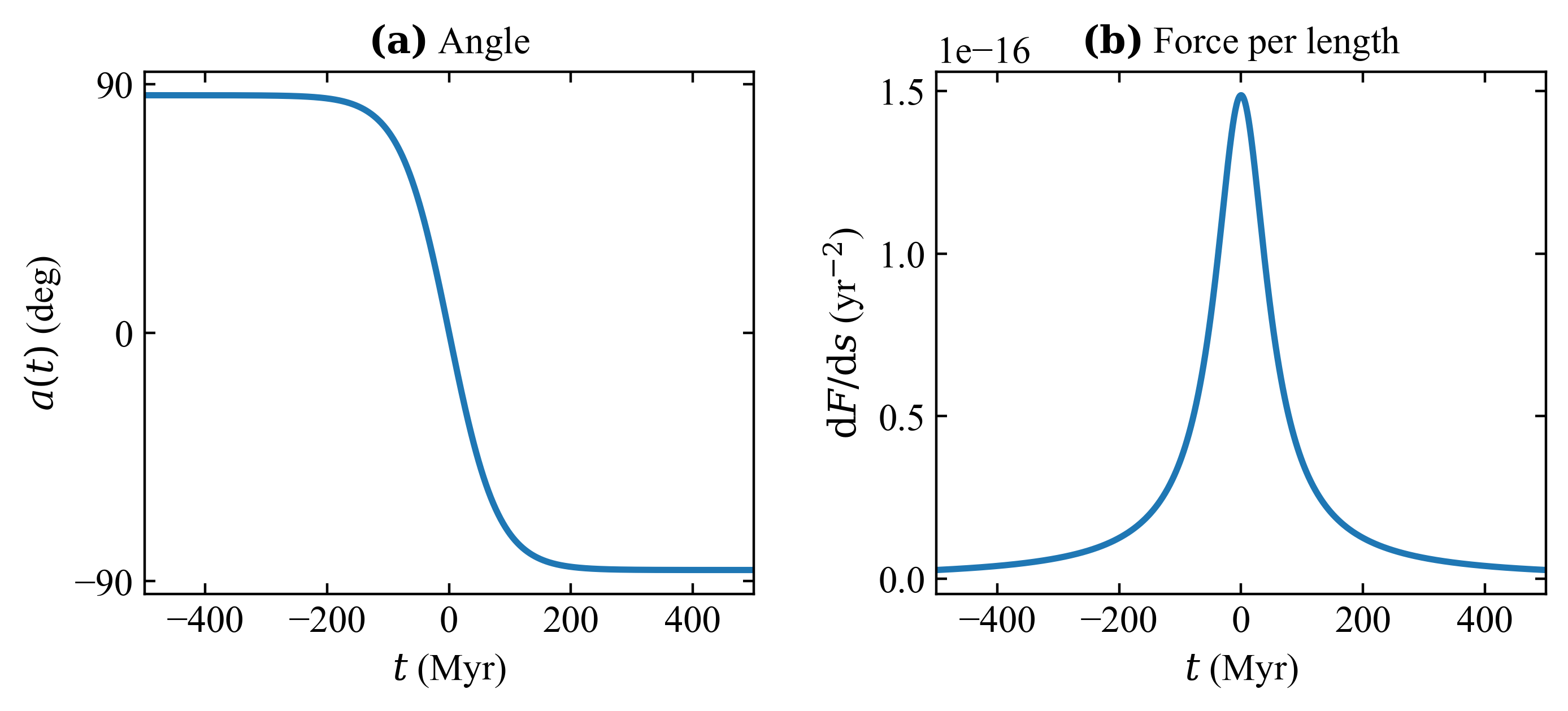}
\caption{Evolution of the tidal force, in the rest frame of the dSph satellite. Panel \textbf{(a)} shows the force's angle, while panel \textbf{(b)} shows the amplitude of its spatial derivative, as defined in eqs.~\eqref{eq:dFds} and \eqref{eq:angle}.
\label{fig:tidal_force}}
\end{figure}

The external tidal force is modelled according to analytic functions, whose numerical values were fitted to the tidal force of an orbit in a Milky Way potential model and then rounded to two significant digits. This is described in detail in appendix~\ref{app:tidal_force}. The tidal gravitational potential takes the form
\begin{equation}\label{eq:tidal_phi}
    \Phi_\text{tidal}(t, \boldsymbol{r}) = 
    \frac{\de F(t)}{\de s}
    \Big\{ -\frac{1}{2} \big[\boldsymbol{r} \cdot \hat{\boldsymbol{r}}_F(t)\big]^2
    + \frac{1}{4} \big|\boldsymbol{r} \times \hat{\boldsymbol{r}}_F(t)\big|^2
    \Big\}.
\end{equation}
In this expression, the quantity
\begin{equation}\label{eq:dFds}
    \frac{\de F(t)}{\de s} = A_T \times 0.12 \, \Bigg( \frac{t^2}{\Myr^2} + 2400 \Bigg)^{-0.86}~\Myr^{-2}
\end{equation}
is the amplitude of the tidal force's spatial derivative, where $A_T$ is a scaling constant close to unity which we vary between our different simulations. The quantity $\hat{\boldsymbol{r}}_F(t)$ is the unit vector in the direction of the external tidal force, equal to
\begin{equation}\label{eq:hatr_F}
	\hat{\boldsymbol{r}}_F(t) = \begin{bmatrix}
	\cos\,a(t) \\
	\sin\,a(t) \\
	0
        \end{bmatrix},
\end{equation}
where
\begin{equation}\label{eq:angle}
    a(t) = -\frac{3}{2} \tanh \left( \frac{t}{80~\Myr} \right).
\end{equation}
The functional forms of eqs.~\eqref{eq:dFds} and \eqref{eq:angle} are illustrated in figure~\ref{fig:tidal_force}. The force amplitude has a full-width at half-maximum of 109~Myr.

\subsection{Fuzzy dark matter simulations}\label{sec:FDM-sim}

In order to study the behaviour of perturbed FDM halos, we use two complementary approaches: on the one hand, we solve the Schr\"odinger-Poisson equation on a grid; on the other hand, we analyse the FDM halo's response using time-dependent perturbation theory. In order to differentiate the two approaches, we refer to the former as a dynamical simulation. They are described in sections~\ref{sec:dynamical_FDM_sim} and \ref{sec:pert_theory} below, respectively.

\subsubsection{Dynamical FDM simulations}\label{sec:dynamical_FDM_sim}

We run dynamical simulations of FDM using the Python package \texttt{PyUltraLight}
\cite{Edwards2018}. This package solves the Schr\"odinger-Poisson equation on a grid, where we choose a spatial resolution of 100~pc. We modify the standard \texttt{PyUltraLight} package by incorporating a time-varying external tidal force, according to section~\ref{sec:tidal_force} above.

\begin{figure}
\centering
\includegraphics[width=1\textwidth]{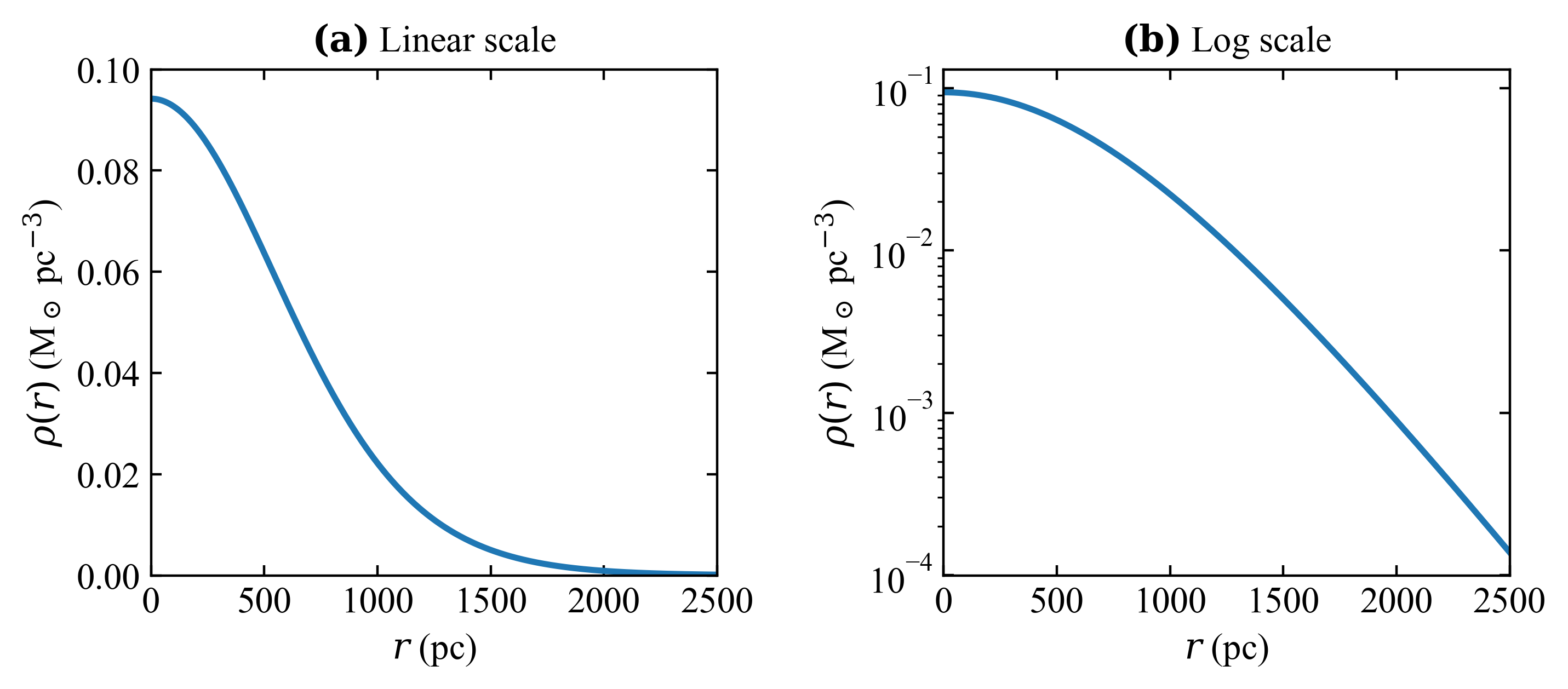}
\caption{Matter density distribution of a soliton with a total mass of $M_\text{sol.} = 3.4 \times 10^8~\Msun$, in linear scale (left panel) and log  scale (right panel). We use this matter density distribution for the initial state of all FDM and CDM simulations presented in this work.}
\label{fig:rho_of_r}
\end{figure}

We initialise the soliton using the standard \texttt{PyUltraLight} initial condition (described in detail in section 3.2 in ref.~\cite{Edwards2018}). We use a soliton mass of $M_\text{sol.} = 3.4 \times 10^8~\Msun$, which has a central density of $\rho_0 = 0.094~\Msunppcc$. The matter density distribution is shown in figure~\ref{fig:rho_of_r}. The characteristic time scale of ground state soliton oscillations is given by \cite{Chiang2021}:
\begin{equation}\label{eq:tsol}
    t_\text{sol.} \simeq 39.9 \left( \frac{\rho_0}{\Msunppcc} \right)^{-1/2}~\Myr,
\end{equation}
which for our initial conditions equals 130~Myr. This is comparable to the time-scale of the tidal force, whose amplitude has a full-width at half-maximum of 109~Myr.

\texttt{PyUltraLight} uses periodic boundary conditions and a Fourier decomposition to solve the Poisson equation. In order to let the FDM wave function escape the box when getting far enough from the central density, every 5~Myr we multiply the wave function by a boundary function.\footnote{This cadence of 5~Myr is short enough such that mass approaching the boundary will escape, while at the same time having a negligible effect on the simulation's inner volume. We also tried other time-steps, as well as different spatial grid sizes and resolutions, but found no appreciable differences when using similar values.} The boundary function is equal to:
\begin{equation}
    B(r) = \text{sigm} \left( \frac{0.45 L - r}{L/100} \right)
\end{equation}
where $\text{sigm}(x) = [1+\exp(-x)]^{-1}$ is the sigmoid function, and $L=30~\kpc$ is the side length of the simulation's spatial grid. This function quickly goes from unity to zero at radii that are close to the half-length of the spatial grid (15~kpc).

Although we only use the fixed mass values described above, the conclusions drawn in this paper are also applicable to other soliton and FDM particle masses. Such systems would react in a fully analogous manner, under a transformation that stretches time and rescales the strength of the external tidal force.
If we rescale the total mass of the soliton by a factor $\alpha$, then the FDM density, the time coordinate, the spatial coordinates, and the tidal force's spatial derivative transform according to
\begin{equation}\label{eq:rescale_a}
    \left\{M,\, \rho,\, t,\, \boldsymbol{r},\, \frac{\de F}{\de s} \right\} \rightarrow
    \left\{\alpha M,\,  \alpha^4\rho,\,  \alpha^{-2}t,\,  \alpha^{-1}\boldsymbol{r},\,  \alpha^4\frac{\de F}{\de s} \right\}.
\end{equation}
If we rescale the FDM particle mass ($m_a$) by a factor $\beta$, while keeping the spatial size of the soliton constant, the same quantities transform like
\begin{equation}\label{eq:rescale_b}
    \left\{m_a,\,  M,\,  \rho,\,  t,\,  \boldsymbol{r},\,  \frac{\de F}{\de s} \right\} \rightarrow
    \left\{\beta m_a,\,  \beta^{-2}M,\,  \beta^{-2}\rho,\,  \beta t,\,  \boldsymbol{r},\,  \beta^{-2}\frac{\de F}{\de s} \right\}.
\end{equation}
The same scaling relations are found in for example Ref.~\cite{Li2021}.

\subsubsection{Time-dependent perturbation theory}\label{sec:pert_theory}

In order to complement and interpret the results from the dynamical simulations, we also analyse the response of the FDM halo using time-dependent perturbation theory. This provides a deeper conceptual understanding as well as support for the results coming from the dynamical simulations. The study of time-dependent perturbations in quantum mechanics is particularly well-suited for this type of problem, in which a ground state wave experiences a perturbing potential.

The technical aspects of the perturbative calculation are described in detail in appendix~\ref{app:perturbation_theory} (see also ref.~\cite{Zagorac2022}). In summary, we perform an expansion of the wave function in the presence of an external, time-dependent perturbation of the potential, where the FDM wave function is constructed as a superposition of eigenmodes in a manner similar to refs.~\cite{Lin2018,Yavetz2022,Zagorac2023}. The FDM halo is initialised to be fully degenerate in its ground-state, and is then partially excited to higher energy modes by the tidal impulse. In order to reduce the complexity of the calculations, we assume that only the amplitudes of the eigenmodes change as a result of the perturbation (but not their energies or shapes). In other words, the library of eigenmodes used in the perturbative calculation is evaluated only once and assuming the initial (unperturbed) potential. As a result, the validity of the results from this calculation decreases for larger tidal perturbations.

\subsection{Massless test-particle simulations}\label{sec:mless-sim}

In order to simulate massless test-particles in an underlying FDM potential, we run $N$-body simulations using the Python package \texttt{REBOUND} \cite{rebound}. The density distribution of the $N$-body particles is set to be equal to the matter density distribution of the corresponding FDM soliton and consists of a total of $5 \times 10^4$ particles. We assume that the satellite's initial conditions are isotropic and in a steady state, which then uniquely determines the full 6d phase-space distribution. We generate the particle velocities using Eddington inversion \cite{Eddington1916,Lacroix2018}.

In the case of the massless test-particles, they are subject to the total gravitational field as given by the sum of the time-varying FDM matter density distribution and the external tidal force. When running the FDM simulations, the total gravitational potential is saved with a temporal and spatial resolution of 5~Myr and 100~pc.

\subsection{Cold dark matter simulations}\label{sec:CDM-sim}

For the CDM case, we also use N-body simulation. These simulations are initialised in the same manner as the massless test-particles, although the CDM particles have a particle mass of $(5\times 10^{4})^{-1} \times 3.4 \times 10^8~\Msun$.

In these simulations, the gravitational potential of the tidal force follows the analytic expressions of eq.~\eqref{eq:tidal_phi}. However, it is implemented into \texttt{REBOUND} as accelerations, given by the spatial first order derivatives of the external gravitational potential.

\section{Results}\label{sec:results}

In this section, we present our results for three simulation setups, labelled A, B, and C, which differ in terms of the tidal field strength. We compare the dynamical simulations of FDM, CDM, and massless test-particles in the FDM gravitational potential. For Simulation A, we also compare the results of the dynamical FDM simulation with those obtained from time-dependent perturbation theory calculations; Simulation A has the weakest tidal force, giving the most accurate perturbation theory prediction (given the assumptions outlined in section~\ref{sec:pert_theory}).

Although we ran many more simulations, we considered these three to be representative of our general results. We saw similar behaviour in other simulations, where we also varied the impulse duration of the tidal force.
Some supplementary results are also shown in appendix~\ref{app:more_results}.

\subsection{Simulation A}
\label{sec:simA}

For this simulation setup, which we label Simulation A, the amplitude of the tidal force, as defined in eqs.~\eqref{eq:tidal_phi}--\eqref{eq:angle}, is re-scaled by a factor of $A_T = 1.4$. With this tidal field strength, the halo loses only a small amount of mass and is induced with lower order excited modes.

\begin{figure}
\centering
\includegraphics[width=\textwidth]{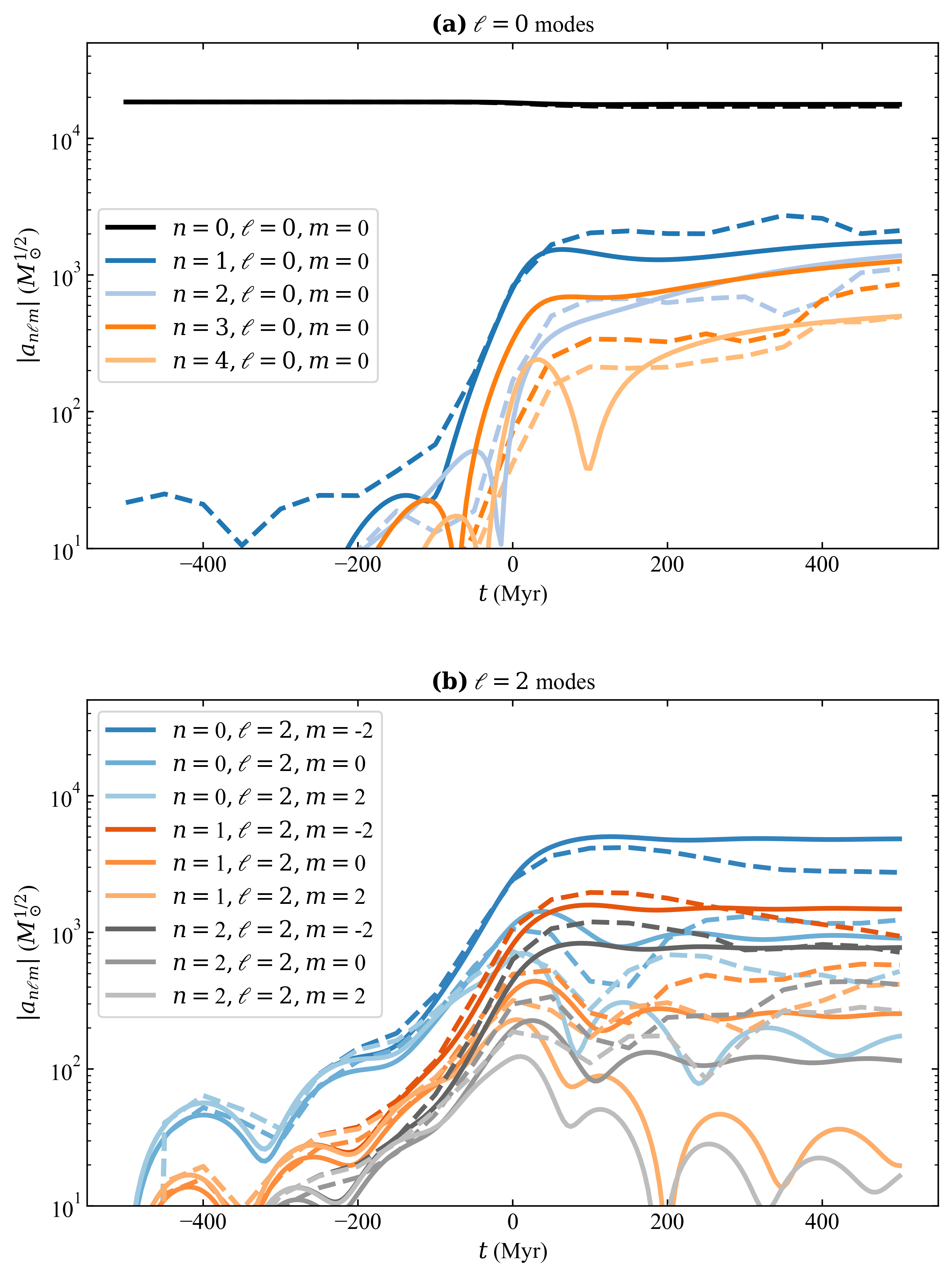}
\caption{Evolution of the absolute value of the eigenmode amplitudes of the FDM halo in Simulation A, based on the time-dependent perturbation theory calculation (solid lines) and the dynamical simulation (dashed lines). The two dominant categories of eigenmodes are depicted in the two panels: in the top panel, the spherically symmetric $\ell=0$ modes; in the bottom panel, the quadrupole $\ell=2$ modes.
\label{fig:pert_theory_amps}}
\end{figure}

We begin with the results from the time-dependent perturbation theory calculations, and compare those with the eigenmode decomposition of our dynamical FDM simulation. As discussed in section~\ref{sec:pert_theory}, the perturbation theory approach is most accurate for a weaker tidal impulse, which is why we apply it specifically to Simulation A. The evolution of the absolute values of the eigenmode amplitudes is shown in figure~\ref{fig:pert_theory_amps}.
The growth of the excited modes comes at the expense of the ground state amplitude, which decreases by roughly 4~\% and 7~\% in the perturbation theory prediction and dynamical simulation, respectively. The dominant excited eigenmodes fall into one of two categories:
\begin{enumerate}
    \item $n\neq0$, $\ell=0$ modes, which are spherically symmetric and cause oscillations of the central density (i.e. a breathing mode);
    \item $\ell=2$, $m=\{-2, 0, 2\}$ modes, which correspond to a rotating quadrupole.
\end{enumerate}
The symmetry of the initial conditions and external force ensures that no modes with odd $\ell$ are excited. Higher order modes (with even values of $\ell$) are also excited, but their contribution to the overall density profile is negligible. A comparison of the amplitudes of the perturbative calculation and the dynamical simulation reveals that they begin to disagree somewhat around the peak of the perturbation ($t=0~\Myr$). At this point in time, the FDM halo loses some mass and changes shape, thus altering the underlying potential and violating the assumptions of the perturbation theory calculation mentioned in section~\ref{sec:pert_theory}. However, the results of the two approaches still agree well, especially for the dominant eigenmodes. After the pericentre passage, when the tidal force has subsided, the perturbation theory predicts static eigenmode amplitudes, such that the FDM halo keeps oscillating indefinitely. Such long-standing modes of oscillations are present also in the dynamical simulations long after the pericentre passage, as is seen in the results presented below. Further results pertaining to the perturbation theory calculation are found in appendix~\ref{app:res_pert_theory}.

\begin{figure}
\centering
\includegraphics[width=0.9\textwidth]{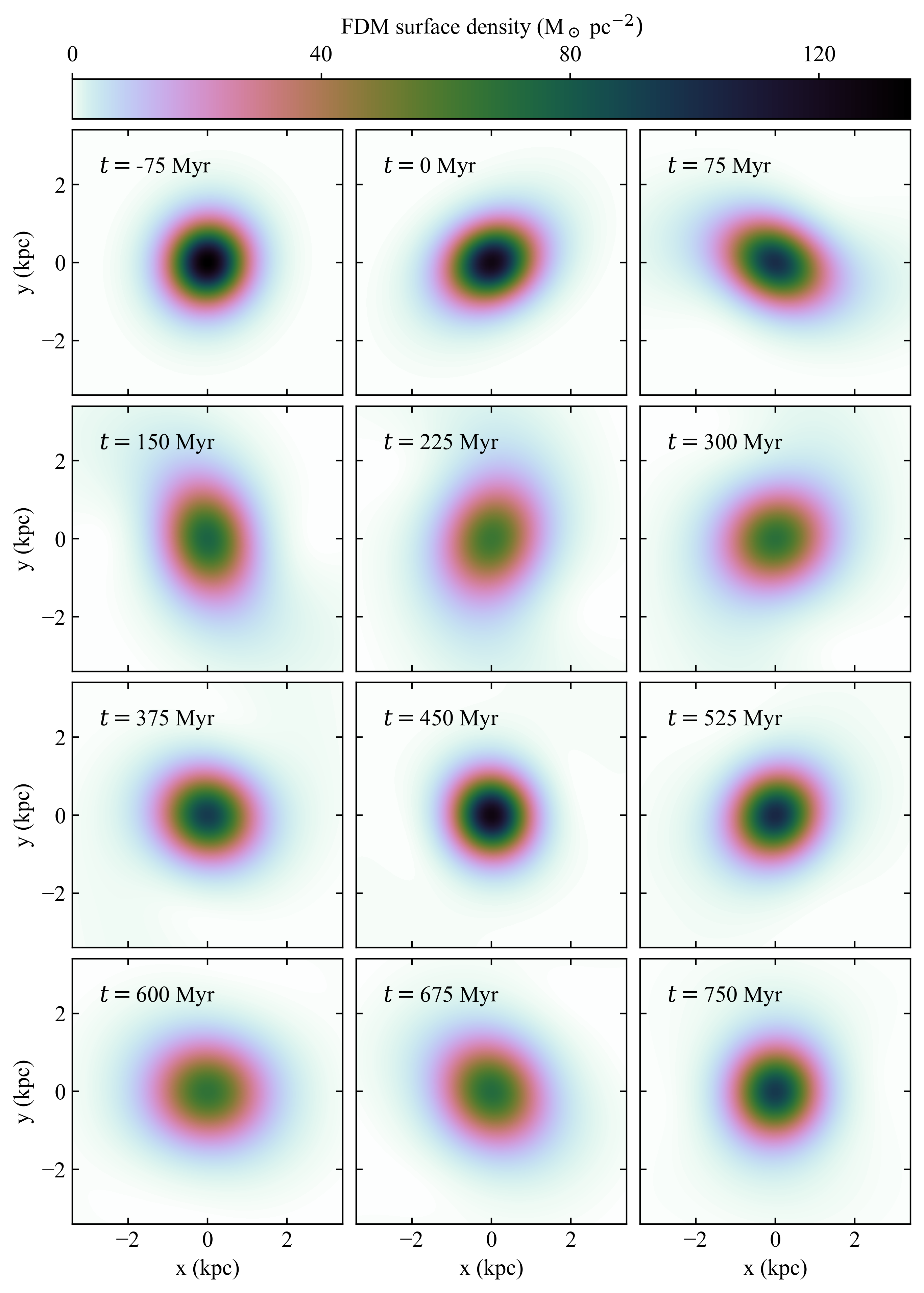}
\caption{FDM surface density in the $(x,y)$-plane for Simulation A, at different snapshots in time, starting at the top left, 75~Myr before the pericentre passage.
\label{fig:FDM_xy_proj_A}}
\end{figure}

In figure~\ref{fig:FDM_xy_proj_A}, we show the dynamically simulated FDM halo in terms of its surface density projected onto the $(x,y)$-plane, for twelve different snapshots between $t=-75~\Myr$ and $t=750~\Myr$. At $t=-75~\Myr$, when the satellite is approaching the pericentre, the soliton is largely undisturbed and only slightly asymmetric from the rising tidal strain. After that, the halo's shape and size vary with time, oscillating long after the tidal force has subsided. This continues in a similar fashion even after $t=750~\Myr$.

\begin{figure}
\centering
\includegraphics[width=1.\textwidth]{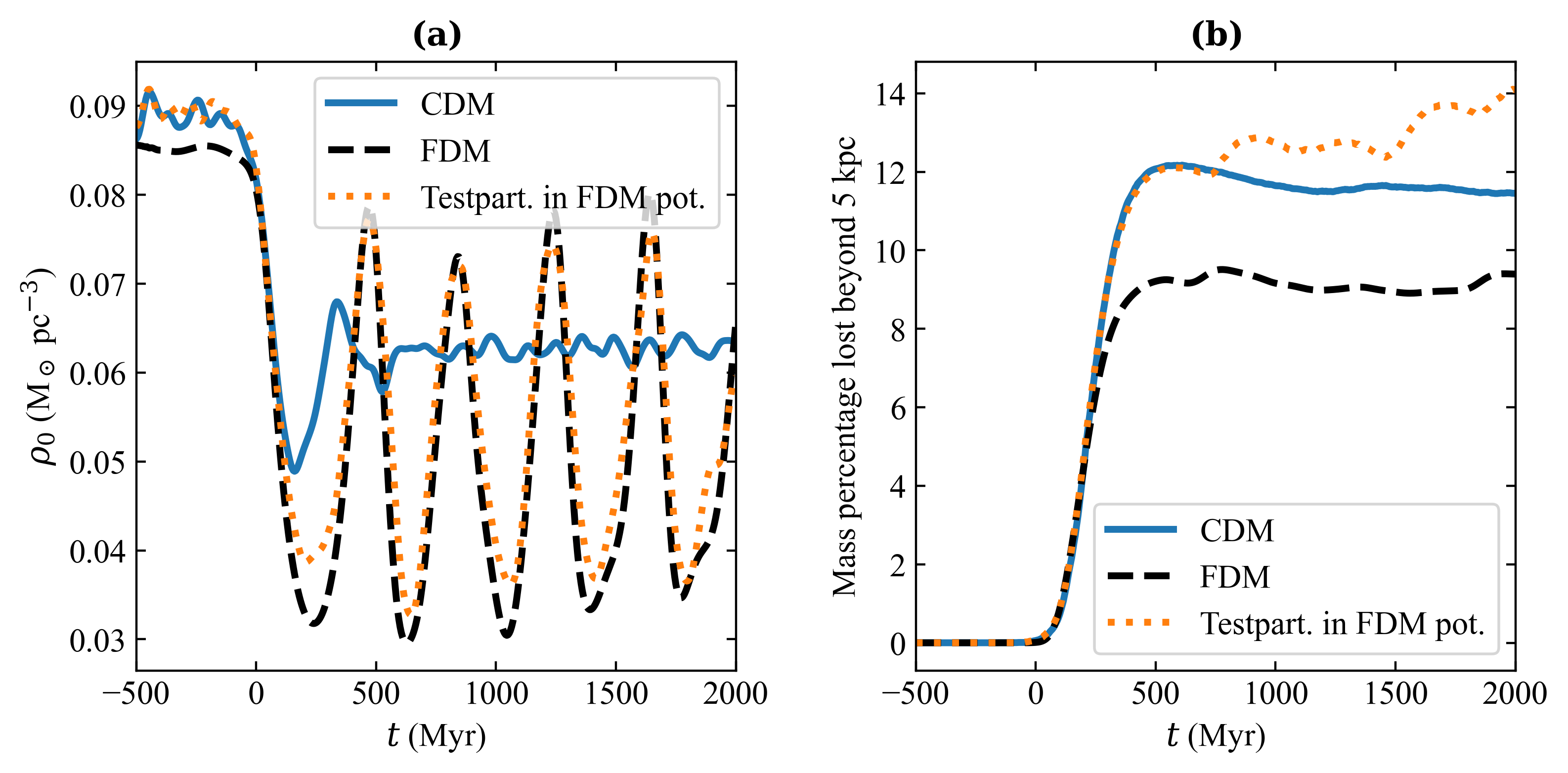}
\caption{Central density ($\rho_0$) and total mass loss of Simulation A, comparing the cases of CDM, FDM, and test-particles in an FDM potential. The test-particles are massless, so the dotted curves are proportional to the test-particles' central number density and total number count. For the CDM and test-particles in panel \textbf{(a)}, the central density is averaged within a spherical volume with a 300~pc radius. The mass loss percentage in panel \textbf{(b)} is given by the mass lost beyond a radius of 5~kpc.
\label{fig:central_density_A}}
\end{figure}

These oscillations are further illustrated in figure~\ref{fig:central_density_A}, showing the evolution of two quantities: the central density and the mass percentage that is lost beyond a radius of 5~kpc. We include all three cases of CDM, FDM, and massless test-particles in the FDM potential. The system's response to the external kick is dramatically different in the CDM and FDM cases. For the CDM, the particles quickly phase-mix and stabilise to a steady state within only a few 100~Myr after the pericentre passage. For the FDM soliton, the central density continues to oscillate long after the pericentre passage, as predicted from the time-dependent perturbation analysis. The FDM oscillations in the simulation are even more pronounced and somewhat slower than the perturbation theory prediction, due to the mass loss and altered shape of the FDM potential. The massless test-particles that inhabit the FDM potential also largely follow the soliton's oscillations. In terms of mass loss, the FDM soliton and the corresponding CDM halo have quite different evolutions, but by $t=2000~\Myr$ they have both lost roughly 12 per cent of their initial mass. The test-particles experience a more significant mass loss fraction, reaching roughly 17 per cent at $t=2000~\Myr$.

\begin{figure}
\centering
\includegraphics[width=1\textwidth]{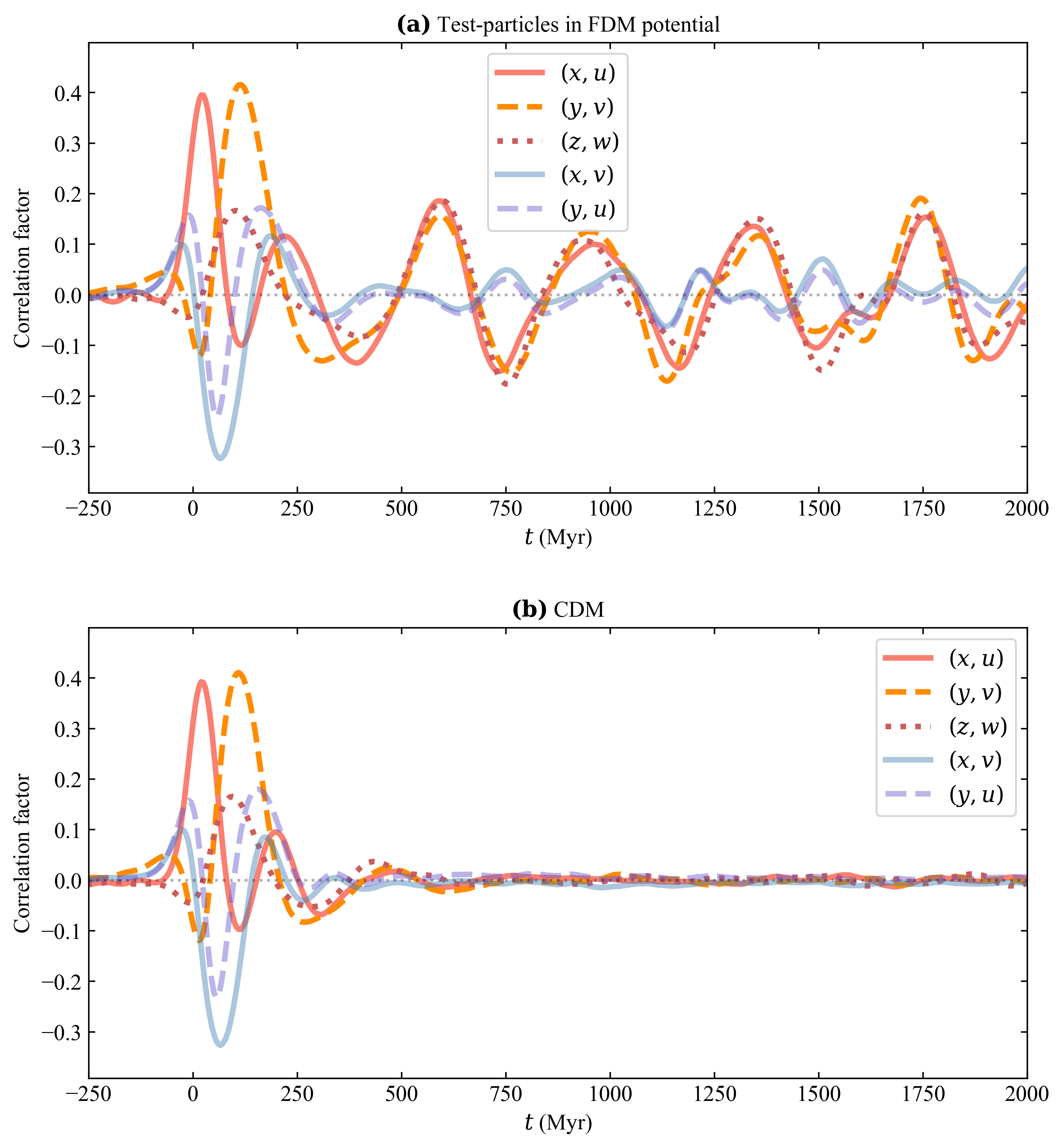}
\caption{Correlations between spatial coordinates and velocities of Simulation A, for particles within a radius of 3~kpc.
Panels \textbf{(a)} and \textbf{(b)} show the test-particles and CDM cases, respectively. Spatial-velocity correlations that are not shown are close to zero at all times. The axis ranges are the same for both panels. The dotted grey line marks a correlation value of zero, for better visibility.
\label{fig:x_vx_correlations_A}}
\end{figure}

In figure~\ref{fig:x_vx_correlations_A}, we show how the correlations between spatial coordinates and velocity coordinates vary with time, both for the massless test-particles in an FDM potential and for CDM particles. Around the time of pericentre, the response is almost identical for the two cases. However, the evolution after the tidal force has subsided ($t \gtrsim 200~\Myr$) is dramatically different. The CDM phase-mixes quickly into a new steady state; this is expected, given that the relaxation time of such a system is on the order of the CDM particles' crossing time
(in this case the crossing time is roughly 100~Myr, see section~\ref{sec:discussion} for further discussion). Conversely, the FDM soliton does not experience such phase-mixing, but instead continues to oscillate. While the test-particles themselves would also phase-mix quickly in a stationary gravitational potential, the FDM potential they inhabit continues to reverberate, dragging the test-particles along.

The non-zero correlation between a spatial coordinate and its time-derivative, such as the $(x,u)$-correlation, is the signature of a distribution that is either expanding or contracting---what is commonly referred to as a breathing mode. If we select test-particles with a lower initial energy, as described at the end of appendix~\ref{app:more_results}, then these correlations become even stronger. The correlations become stronger still if we mask the innermost region by making also a lower cut in radius or projected radius (e.g. by imposing $1<r/\kpc<3$).

The results of this particular simulation, with a persistent breathing mode which also manifests itself on the test-particles, is applicable also in scenarios with weaker tidal perturbations. If we decrease the tidal force amplitude, we see similar structures for CDM, FDM, and test-particle distributions, only less pronounced.

\subsection{Simulation B}
\label{sec:simB}

For this simulation setup, which we label Simulation B, the amplitude of the tidal force is re-scaled by a factor of $A_T = 1.7$. This tidal field is stronger than for Simulation A, but the general characteristics of the dynamical response is very similar.

\begin{figure}
\centering
\includegraphics[width=0.9\textwidth]{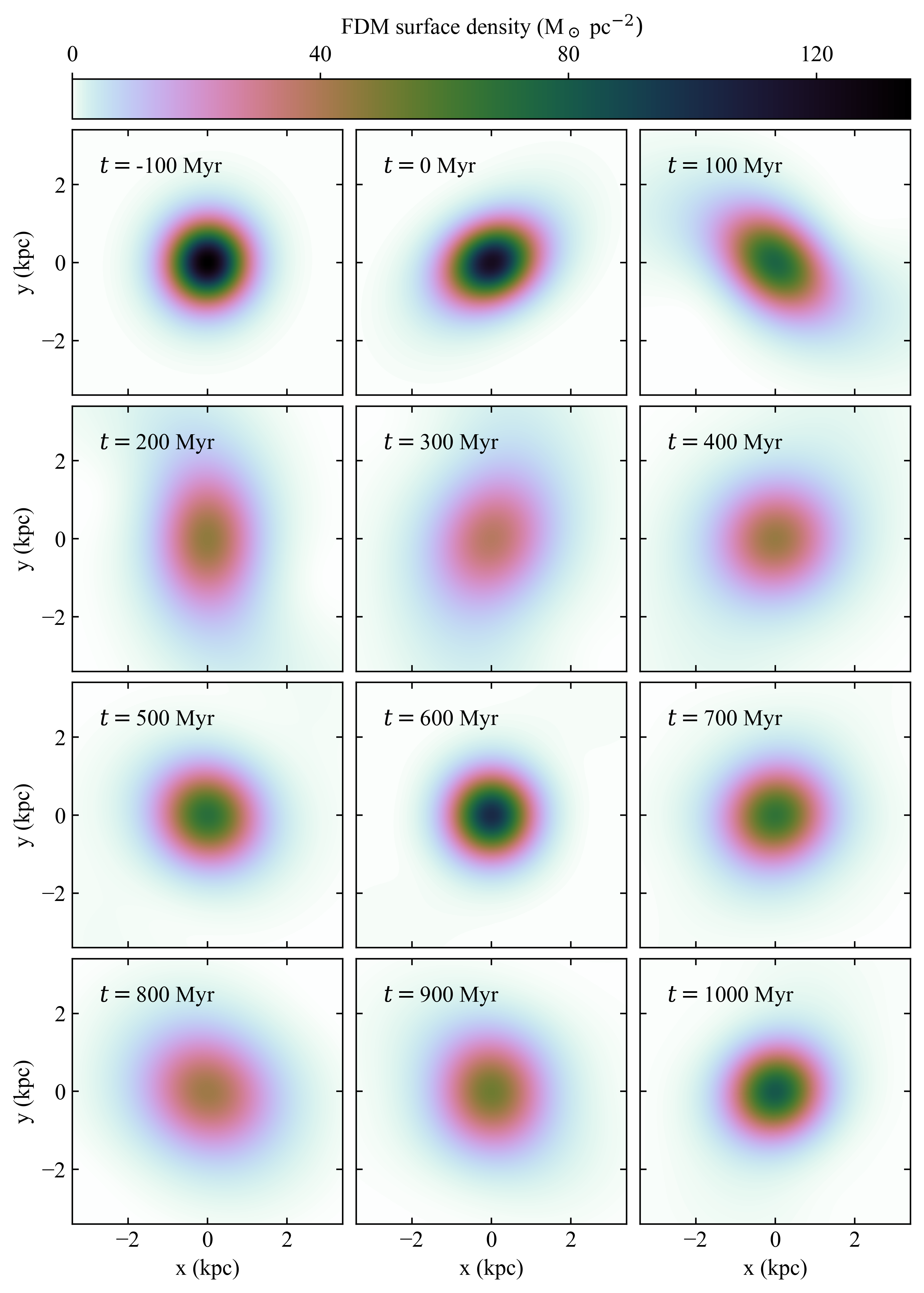}
\caption{Equivalent to figure~\ref{fig:FDM_xy_proj_A}, but for Simulation B, although we note that the time-step between panels is different.
\label{fig:FDM_xy_proj_B}}
\end{figure}

\begin{figure}
\centering
\includegraphics[width=1.\textwidth]{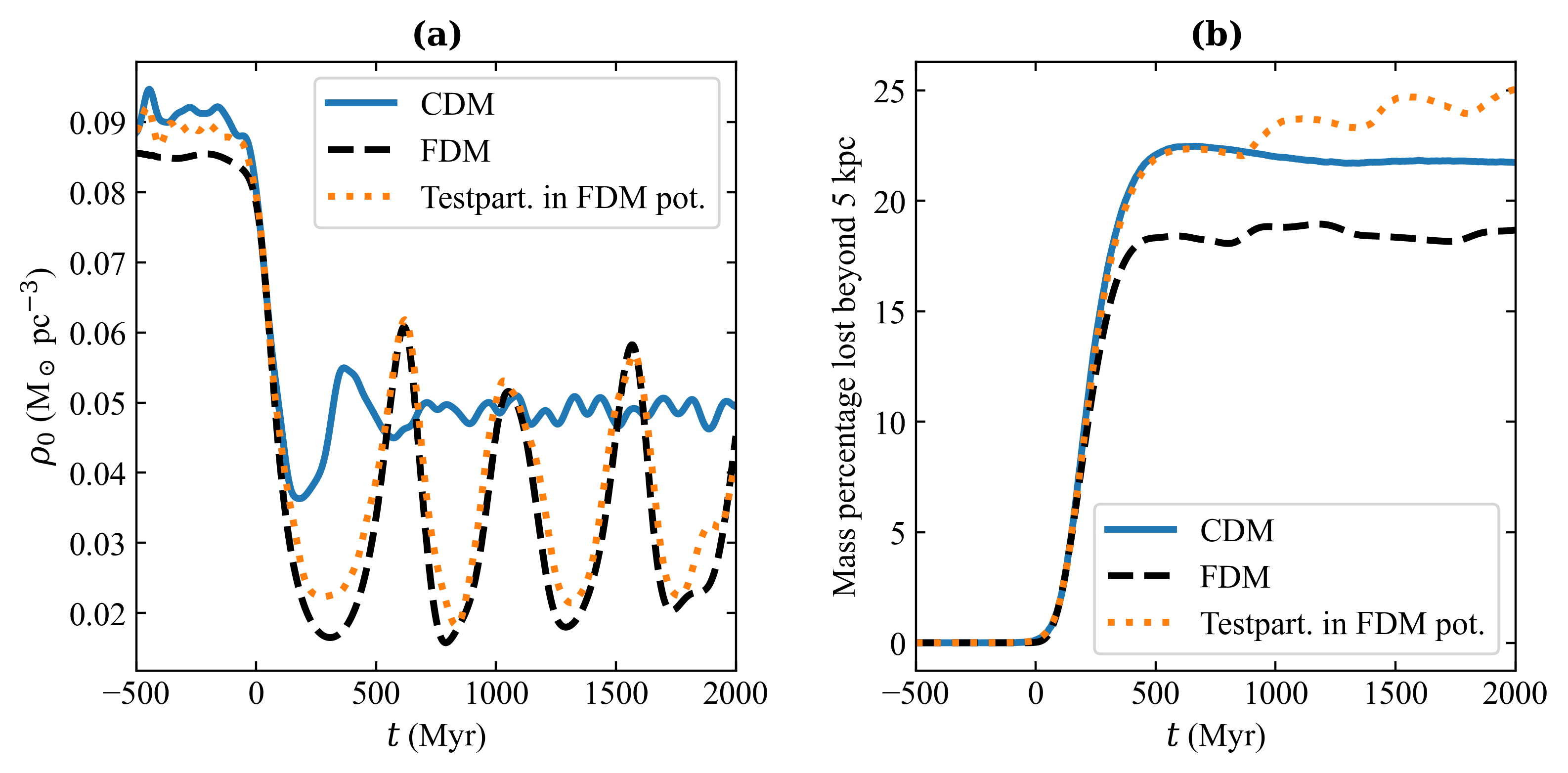}
\caption{Equivalent to figure~\ref{fig:central_density_A}, but for Simulation B, although we note that the extent of the vertical axis in both panels is different.
\label{fig:central_density_B}}
\end{figure}

In figure~\ref{fig:FDM_xy_proj_B}, we show the projected FDM surface density in the $(x,y)$-plane at twelve different times. We see a similar behaviour to Simulation A, as shown in figure~\ref{fig:FDM_xy_proj_A}, although the soliton is more strongly perturbed and reaches lower central densities. This is further illustrated in figure~\ref{fig:central_density_B}. The stronger perturbation strips more mass than for Simulation A (as shown in figure~\ref{fig:central_density_A}), for the CDM, FDM, and test-particles. The CDM particles phase-mix roughly 500~Myr after the pericentre passage, while the FDM and test-particles continue to oscillate, with a slightly longer period than for Simulation A, due to the lower central density of the FDM.

\begin{figure}
\centering
\includegraphics[width=1\textwidth]{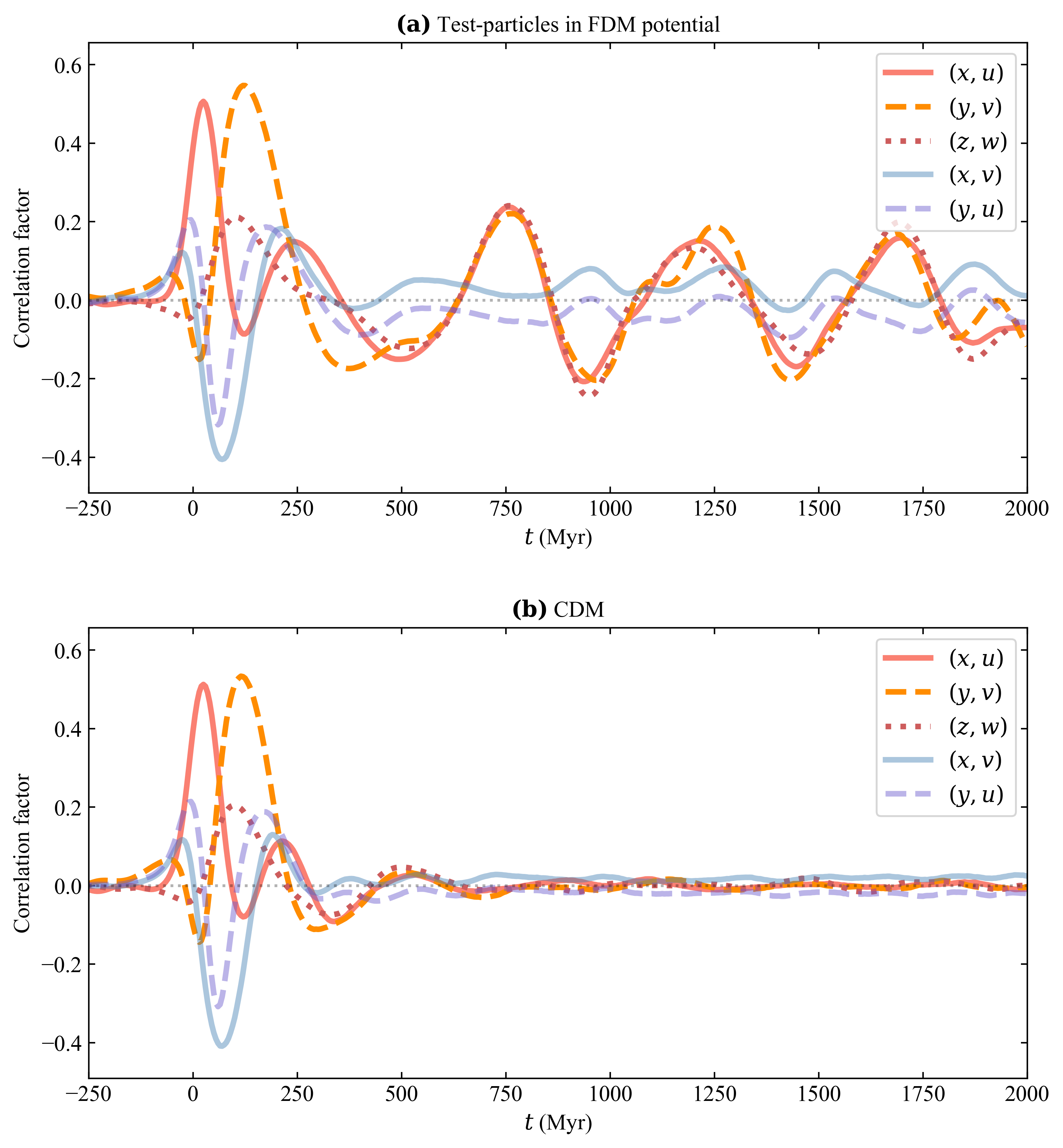}
\caption{Equivalent to figure~\ref{fig:x_vx_correlations_A}, but for Simulation B, although we note that the extent of the vertical axis is different.
\label{fig:x_vx_correlations_B}}
\end{figure}

In figure~\ref{fig:x_vx_correlations_B} we show the
correlation factors between spatial and velocity coordinates, comparing test-particles in an FDM potential to the CDM particles. 
For the test-particles, the correlations
exceed an absolute value of 0.2 long after the pericentre passage, for example around $t\simeq 1700~\Myr$. Conversely, the CDM quickly phase-mixes and stabilises to correlation values close to zero.

In appendix~\ref{app:more_results}, we show further results of Simulation B, mainly in terms of what would be observable if only line-of-sight velocities (but not proper motions) are available. In summary, we see clear differences in the dynamical response of test-particles in an FDM potential and CDM particles, given the chosen initial conditions. For example, the kurtoses of the test-particles spatial distribution vary dramatically, signifying an oscillation between a heavy-tailed, peaked shape and a more box-like shape. However, these differences arise under our specific choice of initial conditions; it is unclear how feasible it is to differentiate FDM and CDM given a greater and more realistic suite of initial conditions. Fully addressing this challenge is beyond the scope of this work, but we discussed this point further in section~\ref{sec:discussion} below.

\subsection{Simulation C}
\label{sec:simC}

For this simulation setup, which we label Simulation C, the tidal force amplitude scaling is set to $A_T = 2$. With this tidal field strength, the FDM soliton is strongly perturbed, and the induced oscillations have a significantly longer period.

\begin{figure}
\centering
\includegraphics[width=0.9\textwidth]{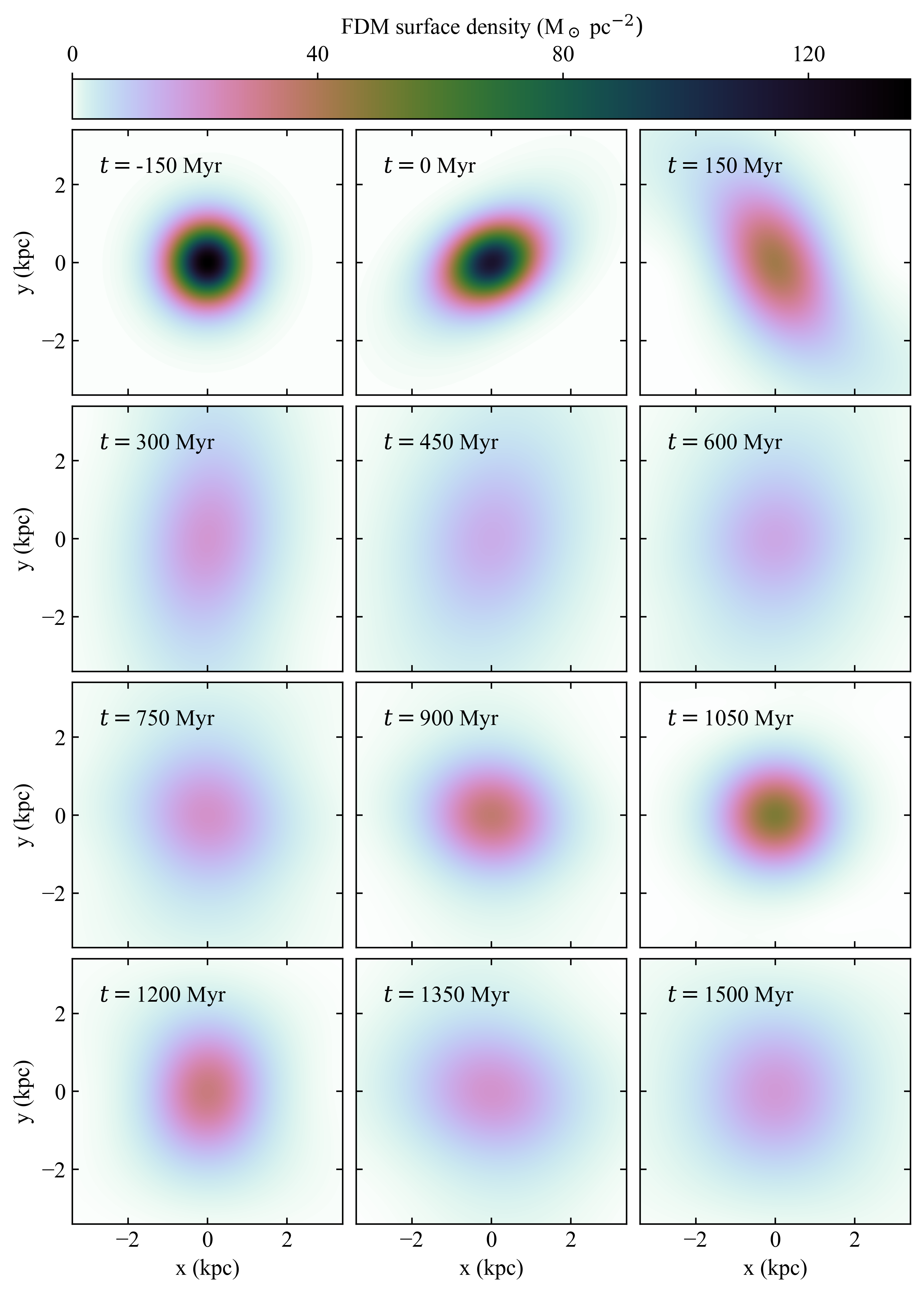}
\caption{Equivalent to figure~\ref{fig:FDM_xy_proj_A}, but for Simulation C, although we note that the time-step between panels is different.
\label{fig:FDM_xy_proj_C}}
\end{figure}

In figure~\ref{fig:FDM_xy_proj_C}, we show the projected FDM surface density in the $(x,y)$-plane at twelve different times. A few 100~Myr after the pericentre passage, it might seem like the FDM soliton is completely destroyed, but most of its initial mass is in fact still held together by self-gravity. At $t \simeq 1000~\Myr$, it is again contracting.

\begin{figure}
\centering
\includegraphics[width=1.\textwidth]{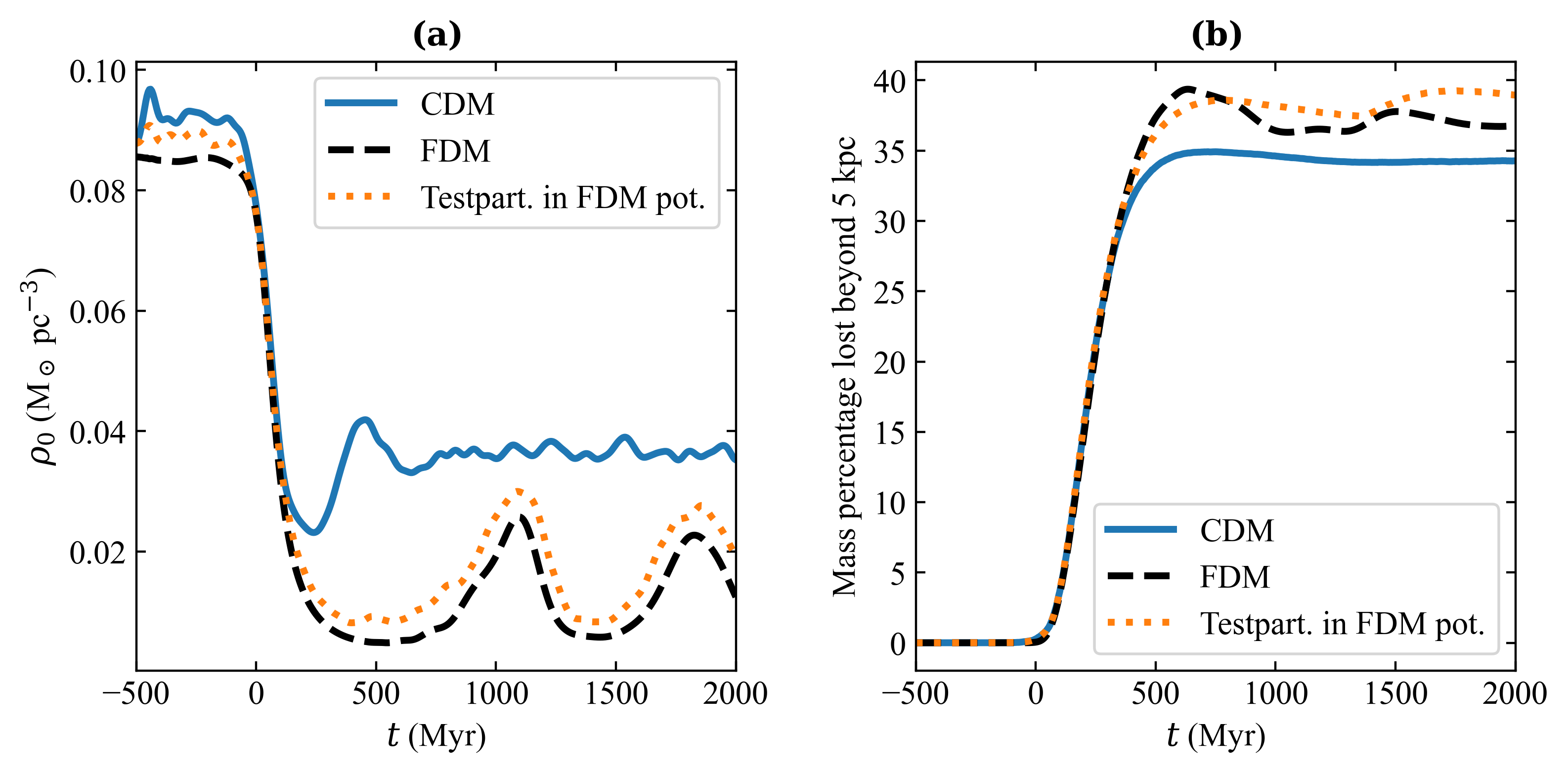}
\caption{Equivalent to figure~\ref{fig:central_density_A}, but for Simulation C, although we note that the extent of the vertical axis in both panels is different.
\label{fig:central_density_C}}
\end{figure}

\begin{figure}
\centering
\includegraphics[width=1\textwidth]{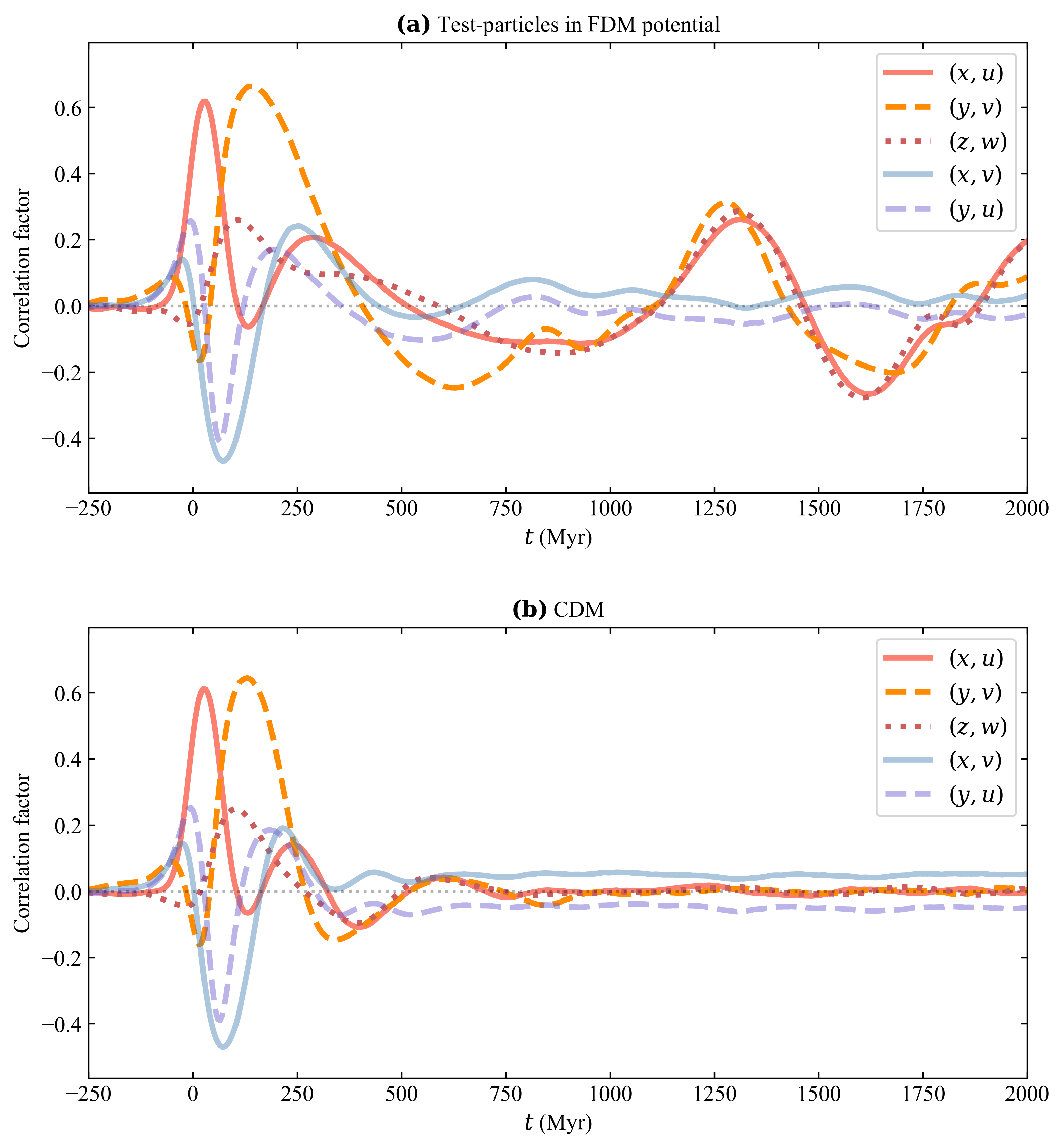}
\caption{Equivalent to figure~\ref{fig:x_vx_correlations_A}, but for Simulation C, although we note that the extent of the vertical axis is different.
\label{fig:x_vx_correlations_C}}
\end{figure}

Its oscillations are further illustrated in panel \textbf{(a)} of figure~\ref{fig:central_density_C}. The FDM distribution reaches central density values extremely close to zero, but has sharp peaks with a periodicity of almost 1000~Myr. Both the CDM and FDM distributions lose roughly a third of their initial total mass. The test-particles in the FDM potential similarly lose marginally more. An interesting feature is that the test-particle mass loss has a step-wise behaviour, as a result of the convulsions of the underlying FDM distribution. This creates a delayed expulsion which does not have a counterpart in the CDM case.

In figure~\ref{fig:x_vx_correlations_C} we show the correlation factors between spatial and velocity coordinates. For the test-particles, the correlations
exceed an absolute value of 0.3 more than a billion years after pericentre passage, while the CDM quickly phase-mixes within a few hundreds of Myr. Even in this rather extreme case, where the tidal force has stripped a large amount of the soliton mass, mainly along the $y$-axis, the FDM halo and associated test-particles return to a state of spherical symmetry at roughly $t\simeq 1~\Gyr$, after which the phase-space correlations evolve synchronously in all three dimensions.

\section{Discussion}\label{sec:discussion}

In this work, we have studied the internal dynamics of a dSph on an eccentric orbit around its host galaxy, which is tidally perturbed during its pericentre passage. We have compared the response of an FDM soliton with the corresponding cored CDM halo and have found fundamental differences in how they phase-mix and dissipate. These differences also imprint themselves on a dSph's stellar population.

We find that the most general and prominent FDM signature is a persistent breathing mode, caused by interference between the ground state and the first few excited states of the FDM wave function. Although there are some non-linear gravitational self-interactions between the modes, the FDM does not phase-mix in the same manner as a CDM halo, given that the former consists of only a small number of specific energy-momentum states. A CDM halo phase-mixes in a short time, washing out the tidally induced breathing mode on time-scales comparable to the dSph's crossing time. Not only do the respective DM distributions react very differently, we also see clearly that a stellar population inhabiting the FDM soliton would be dragged along in these breathing oscillations (e.g. figures~\ref{fig:central_density_A}, \ref{fig:central_density_B}, and \ref{fig:central_density_C}). Such a breathing mode can be detected by positive or negative correlations between a spatial coordinate and the velocity in the same direction (e.g. $x$ and $u$).

In terms of measuring and resolving the internal velocity field of a dSph, the line-of-sight velocity is the easiest component to obtain. Unfortunately, it is unfeasible to resolve a dSph's spatial extent along the line-of-sight (e.g. via parallax); hence, the distance-$v_\text{l.o.s}$ pair is not available. Instead, one would have to rely on sky angle coordinates and proper motion measurements in order to detect a correlation between a spatial coordinate and its corresponding velocity. As discussed in section~\ref{sec:intro}, \emph{Gaia}'s proper motion measurements are starting to resolve the internal dSph dynamics of some objects, which will improve further with future data releases. In fact, the precision of proper motion measurements improves significantly faster than for the parallax (the uncertainties scale like $T^{-3/2}$ and
$T^{-1/2}$, respectively, where $T$ is the total observation time \citep{GaiaEDR3}). The upcoming LSST survey will provide proper motions and significantly greater depths than the \emph{Gaia} liming magnitude. Furthermore, the combination of surveys can produce extremely precise proper motion measurements. Ref.~\cite{GaiaHSTcombined2022} discusses how data from the Hubble Space Telescope and \emph{Gaia} DR5 could produce proper motions with a precision of a few $10^{-2}~\masyr$, even for stars close to \emph{Gaia}'s apparent magnitude limit. For a dSph at a distance of 100~kpc, $10^{-2}~\masyr$ corresponds to a velocity precision of $4.5~\kmsec$, which would likely be enough to resolve a dSph's internal kinematics given a larger amount of statistics. Given these near future advances, it could be possible to detect the signatures identified in this work, if present in dSphs around our own Milky Way.

Although acquiring sufficiently precise proper motion observations to resolve the internal kinematics of dSphs is feasible, it is much easier to measure precise line-of-sight velocities. In appendix~\ref{app:more_results}, we showed what could be observable when limited to this single velocity component. Given the chosen initial conditions, we do see clear differences between the simulated FDM and CDM cases. For example, the test-particles in the FDM potential have phase-space kurtoses that oscillate dramatically. Furthermore, the CDM halo re-forms its core and returns to a new stable central density and spatial configuration significantly faster than the corresponding FDM soliton. However, it is unclear to what extent these differences are reconcilable if given a greater freedom in choosing the initial conditions of our simulations, especially when observations are limited to fewer phase-space dimensions. Potentially, the dynamical properties of FDM and CDM could still manifest themselves in an observable manner, even when limited to velocities along the line-of-sight. This would require the development of a more sophisticated method of inference, which we leave for future work.

As discussed above, we have made some simplifying assumptions concerning the simulations' initial conditions and the perturbing tidal force, but despite this, our result is general. For example, the CDM density profile was assumed to be perfectly proportional to that of the FDM soliton. However, having a cuspy profile would not help reproduce the breathing mode signature seen in our FDM simulations or otherwise change our conclusions; in fact, a cuspy profile would phase-mix and quench a breathing mode even faster than a cored CDM profile. We also tested different initial distributions of test-particles, by making cuts in initial energy and imposing anisotropy (described in greater detail at the end of appendix~\ref{app:more_results}), but saw similar results. Changing the precise properties of the tidal force (e.g. the impulse duration) also did not alter our main result. Similarly, our conclusions are applicable also to other FDM particle masses; such FDM solitons are completely analogous if rescaling the time and the strength of the tidal force.

\section{Conclusion}\label{sec:conclusion}

In this work we have used perturbation theory and simulations to study the dynamical signatures of FDM compared to CDM in tidally perturbed dSphs, in the idealised but still realistic scenario of a fully degenerate FDM soliton which is perturbed by a short-lived external impulse from tidal forces. We have focused on time-varying dynamical signatures that are strong and easily identifiable, even by eye inspection of phase-space histograms and correlation plots. Our main finding is that a perturbed FDM soliton, unlike a CDM halo, develops a long lived breathing mode, whose oscillations would also imprint themselves on an observable stellar population.

Such dynamical signals could be observable with near future instruments. Having proper motion measurements that are precise enough to resolve the inner velocity field of a dSph is especially useful, if not necessary in most cases, in order to detect the long lived breathing that we found to be characteristic of an FDM soliton. Such proper motion precision could be achievable, for example, by combining different survey observations with a large separation in time (e.g. \cite{GaiaHSTcombined2022}).

This work shows that the stellar dynamics of dSphs are informative of the dynamical properties of DM and, by extension, its particle nature. It demonstrates that non-equilibrium dynamics can be particularly informative, allowing for the extraction of information that would not be available in an equilibrium scenario. It further motivates similar studies in the context of other DM models, such as self-interacting or superfluid dark matter, which also have dynamical properties that could manifest on the spatial and temporal scales relevant to dSphs.

\acknowledgments

AW is supported by the Carlsberg Foundation via a Semper Ardens grant (CF15-0384). TDY is supported through the Corning Glass Works Foundation Fellowship at the Institute for Advanced Study. XL is supported by NSERC, funding reference \#CITA 490888-16 and the Jeffrey L. Bishop Fellowship.




\bibliographystyle{JHEP}
\bibliography{biblio.bib}


\appendix

\section{External tidal force}\label{app:tidal_force}

In this section, we demonstrate that the tidal force we apply is realistic, given a Milky Way-like gravitational potential and a dSph on an eccentric orbit. We use the gravitational potential from ref.~\cite{Bovy2015}, implemented as the standard \texttt{MilkyWayPotential} in the \texttt{gala} Python package \cite{PriceWhelan2017}, which consists of a spherical nucleus and bulge, a Miyamoto-Nagai disk, and a spherical NFW dark matter halo.

\begin{figure}
\centering
\includegraphics[width=1.\textwidth]{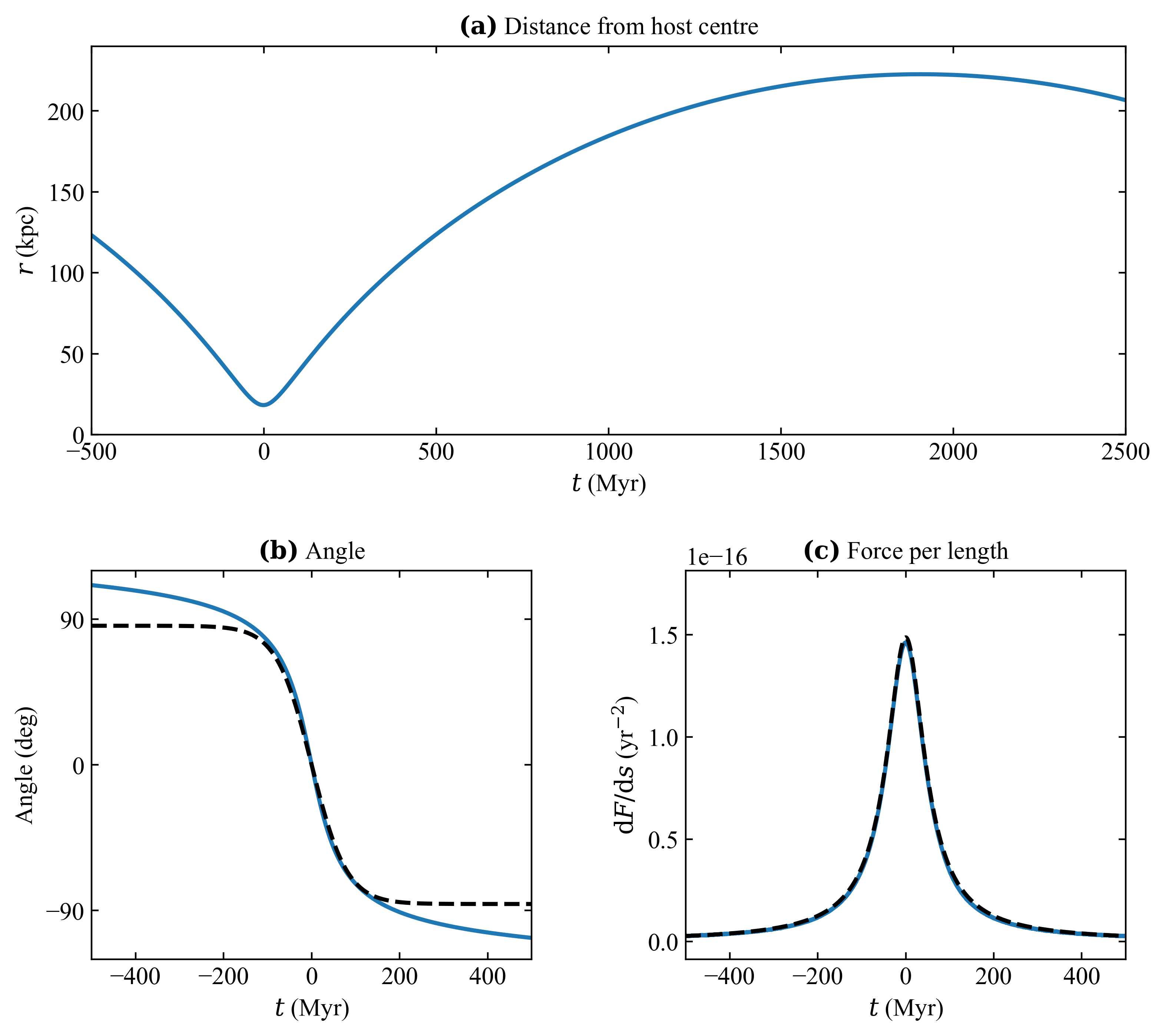}
\caption{Tidal force acting on an object on an eccentric orbit in a Milky Way-like gravitational potential.
The dashed black lines in the bottom panels represent the analytic form that we use in our simulations, as described in section~\ref{sec:tidal_force}. See the main text for further details.}
\label{fig:tidal_force_4appendix}
\end{figure}

We start by simulating the orbit of a single particle, with initial phase-space coordinates $\boldsymbol{X} = (129,0,0)~\kpc$ and $\boldsymbol{V} = (-152,0,59)~\kmsec$, as given in the host galaxy's rest frame. The evolution of the distance between the satellite particle and the host galaxy's centre (labelled $r$) is shown in the top panel of figure~\ref{fig:tidal_force_4appendix}. The satellite particle reaches a minimum radius of 18.25~kpc. This is comparable to, for example, Crater II which is on an eccentric orbit with a pericentre radius of $24.0^{+5.6}_{-5.2}~\kpc$ \cite{Pace2022}.

We calculate the tidal force acting on the simulated particle, written $\de F / \de s$, by taking a numerical derivative in the direction of the host galaxy's centre ($\hat{r}_F$), according to
\begin{equation}
    \frac{\de F}{\de s} = \frac{
    F_{r} [\boldsymbol{X}+(1~\kpc) \times \hat{\boldsymbol{r}}_F] -
    F_{r} [\boldsymbol{X}-(1~\kpc) \times \hat{\boldsymbol{r}}_F]
    }{2~\kpc}.
\end{equation}
We then perform an analytic fit, using the simple functions described in section~\ref{sec:tidal_force}, and round the fitted parameter values to two significant digits. A comparison between the numerical values of the simulation and the analytic function is shown in the two bottom panels of figure~\ref{fig:tidal_force_4appendix}. In panel \textbf{(b)}, showing the angle of the tidal force, the dashed black line does not align perfectly with the simulation. However, they do agree well in the region where the tidal force has a significant amplitude ($|t|\lesssim 200~\Myr$). A more complicated function could better model the angle also at greater values for $|t|$,
but since it would not have any impact on the conclusions of this work, we opted for the simple functional form of eq.~\eqref{eq:angle}.

The orbit used here is meant to be realistic, but not necessarily to liken any specific Milky Way dSph satellite. We take the liberty to rescale the tidal force by the constant $A_T$, see eq.~\eqref{eq:dFds}. The conclusions drawn in this work are very general and apply also to other orbits with, for example, varying degrees of eccentricity or tidal field strengths. The strength of the dSph's response is very sensitive to its total mass, as described at the end of sect.~\ref{sec:dynamical_FDM_sim}, making it possible to have an analogous dynamical response also in a scenario with a significantly weaker tidal force.

\section{Time-Dependent Perturbation theory}\label{app:perturbation_theory}

In this appendix, we describe the general application of time-dependent perturbation theory in this work, and how we make use of it to understand the behaviour of a perturbed dwarf galaxy that is initially dominated by the FDM ground state.

An FDM halo in equilibrium is well-modeled as a wave function ($\Psi$) that satisfies the time-dependent Schr{\"o}dinger equation:
\begin{equation}
    \label{eq:time_dependent_schrodinger}
    i\hbar\frac{\partial \Psi}{\partial t} = H\Psi \ ,
\end{equation}
in which we begin by considering the unperturbed Hamiltonian $H$:
\begin{equation}
    \label{eq:unperturbed_H}
    H = H^0 \equiv -\frac{\hbar^2}{2m_a}\nabla^2 + m_a\Phi_0 \ .
\end{equation}

Because the unperturbed potential (denoted by $\Phi_0$) is fixed and spherically symmetric, the time-dependent wave function can be constructed as a superposition of eigenmodes with fixed amplitudes and energies:
\begin{equation}
    \label{eq:wavefunction}
    \Psi(\bm{x}, t) = \sum_j a_j \psi_j e^{-iE_jt / \hbar} \ ,
\end{equation}
in which $\psi_j$ represents the $j$th eigenmode, and $a_j$ and $E_j$ are its associated amplitude and energy, respectively. In turn, each eigenmode can be written as the product of a radial component and a spherical harmonic:
\begin{equation}
    \label{eq:eigenmode}
    \psi_{n\ell m} = R_{n\ell}(r)Y_\ell^m(\theta, \phi) \ ,
\end{equation}
where we have replaced $j$ with $n$, $\ell$, and $m$, denoting the radial, azimuthal, and magnetic quantum numbers, respectively. The $Y_\ell^m$ term represents the standard spherical harmonic functions, and the radial term $R_{n\ell}$ satisfies the radial part of the time-independent Schr{\"o}dinger equation:
\begin{equation}
    \label{eq:time_independent_schrodinger}
    E_{n\ell}R_{n\ell} = -\frac{\hbar^2}{2m_ar^2}\frac{d}{dr}\bigg(r^2\frac{dR_{n\ell}}{dr}\bigg) + \bigg[\frac{\hbar^2}{2m_ar^2}\ell(\ell+1) + m_a\Phi_0\bigg]R_{n\ell} \ .
\end{equation}

Assuming the FDM is itself the source of the unperturbed gravitational potential, and because the total number density of particles is $|\Psi|^2$, self-consistency requires the wave function to satisfy the Poisson equation:
\begin{equation}
    \label{eq:Poisson}
    \nabla^2\Phi_0 = 4\pi G\rho = 4\pi Gm_a|\Psi|^2 \ .
\end{equation}

Up to this point, we have dealt with a static setup involving a time-independent potential.\footnote{It is worth noting that even in this ``time-independent'' setup, the wave function fluctuates as a result of interference between the different eigenmodes in eq.~\eqref{eq:wavefunction}, leading to time-dependent changes in the potential. However, constructing FDM haloes based on the time-averaged, static gravitational potential accurately reflects the behaviour of FDM haloes in dynamical simulations \cite{Yavetz2022}.}
Of course, the main focus of this work is to study the effect of tidal perturbations on FDM halos, so we now proceed to describe the time-dependent perturbation theory we use to estimate the effect of introducing a small tidal perturbation such as the one described in section~\ref{sec:tidal_force}. Our goal is to understand how $\Psi$ changes when we replace $H = H^0$ in eq.~\eqref{eq:time_dependent_schrodinger} with:

\begin{equation}
    \label{eq:perturbed_H}
    H = H^0 + H' = H^0 + m_a\Phi_\mathrm{tidal}(t) \ .
\end{equation}

For small perturbations, we assume the resulting change in the overall wave function is small enough to avoid causing a significant change in $\Phi_0$. This allows us to hold the eigenmodes $\psi_j$ and eigenvalues $E_j$ constant, and the only difference is the time-dependence of the amplitudes $a_j(t)$ in eq.~\eqref{eq:wavefunction}. This assumption is the main source of any discrepancies between the perturbation theory calculations and the simulations. In reality, eq.~\eqref{eq:Poisson} stipulates that any changes to $\Psi$ will necessarily change $\Phi_0$, causing changes to the library of eigenmodes calculated using eq.~\eqref{eq:time_independent_schrodinger}. A more accurate calculation would involve updating the potential $\Phi_0$ at every step based on the evolving wave function (e.g. ref.~\cite{Zagorac2022}). However, for the purposes of this work, we find the results from this simplified version of the calculation to be sufficient for describing the qualitative behaviour of the simulations.

We insert the revised forms of eq.~\eqref{eq:wavefunction} into the perturbed eq.~\eqref{eq:time_dependent_schrodinger}, and leverage the time-independent solutions to simplify the resulting equation:
\begin{equation}
    \label{eq:pert_theory1}
    i\hbar\sum_j \dot{a}_j(t)\psi_je^{-iE_jt / \hbar} = \sum_j a_j(t)[H'\psi_j]e^{-iE_jt / \hbar} \ .
\end{equation}

Next, we take the inner product of eq.~\eqref{eq:pert_theory1} with $\psi_i$, and leverage the orthogonality of the eigenmodes  $\langle \psi_j | \psi_i \rangle = \delta_{ij}$, to obtain:

\begin{equation}
    \label{eq:pert_theory2}
    i\hbar\big(\dot{a}_i(t)e^{-iE_it/\hbar}\big) = a_i(t) H'_{ii} + \sum_{j\neq i} a_j(t) H'_{ij}e^{-iE_jt\hbar} \ ,
\end{equation}
where we define $H'_{ab} \equiv \langle \psi_a | H' | \psi_b \rangle $ and note that $H'_{ii} = 0$. Solving for $\dot{a}_i(t)$ yields:
\begin{equation}
    \label{eq:pert_theory3}
    \dot{a}_i(t) = -\frac{i}{\hbar}\sum_{j\neq i}a_j(t)H'_{ij}e^{-i(E_j - E_i)t / \hbar} \ .
\end{equation}

To zeroth order (equivalent to the situation with no perturbation at all), the amplitudes remain constant at their initial value at $t=0$: $a_j^{(0)}(t) = a_j(0)$ (we use the superscript to denote the order of the approximation). The first order approximation can now be obtained by inserting the zeroth order solution into eq.~\eqref{eq:pert_theory3}:
\begin{equation}
    \label{eq:1st_order_dt}
    \dot{a}_i^{(1)} = -\frac{i}{\hbar}\sum_{j\neq i} a_j^{(0)}(t) H'_{ij}e^{-i(E_j - E_i)t / \hbar}\ ,
\end{equation}
\[\Downarrow\]
\begin{equation}
    \label{eq:1st_order}
    a_i^{(1)}(t) = a_i(t=0) - \frac{i}{\hbar}\sum_{j\neq i}\int_0^t a_j^{(0)}(t')H'_{ij}e^{-i(E_j - E_i)t'/\hbar}dt' \ .
\end{equation}

We repeat this process once more to obtain the second order approximation, this time inserting the first order solution from eq.~\eqref{eq:1st_order} into eq.~\eqref{eq:pert_theory3}. This will prove necessary for obtaining the time evolution of several eigenmodes, including the time-dependence of the ground state amplitude.

In order to match the dynamical simulations, we set the FDM particle mass to $m_a = 10^{-22}~\eV$ and construct a soliton (i.e. a wave function consisting of only the ground state eigenmode) with a total mass of $M_\mathrm{sol.} = 3.4 \times 10^8~\Msun$, by solving eqs.~\eqref{eq:time_independent_schrodinger} and \eqref{eq:Poisson} (see also ref.~\cite{Hui2017}). The initial amplitudes are thus:
\begin{equation}
    a_j(t=0) = 
    \begin{cases}
        \sqrt{M_\mathrm{sol.}} & \text{for}\quad j = 0 \ , \\
        0 & \text{for}\quad j \neq 0 \ .
    \end{cases}
\end{equation}

Combining eq.~\eqref{eq:1st_order} with these initial conditions allows us to calculate the evolving amplitudes of the various FDM eigenmodes in the presence of the time-dependent perturbing potential $\Psi_\mathrm{tidal}(t)$. We also note that given these initial conditions, several of the eigenmodes (specifically, all the $\ell=0$ modes, including the ground state) have $\dot{a}^{(1)}(t) = 0$, and so it is necessary to compute the second order approximation to obtain the time evolution of their amplitudes.

To study the evolution of the wave function eigenmodes in the dynamical simulations, we again take advantage of the orthogonality of the individual eigenmodes to decompose the halo and compute the respective amplitudes, according to
\begin{equation}
    \label{eq:decompose}
    \langle \Psi | \psi_i \rangle = \bigg\langle{\sum_j a_j \psi_j e^{-iE_jt / \hbar} \bigg| \psi_i \bigg\rangle = \sum_j \delta_{ij} a_j} = a_i \ .
\end{equation}
We compute the eigenmode amplitudes for each timestep, accounting for the spherically averaged potential at that time. These amplitudes are useful for understanding how the halo has evolved away from a pure ground state soliton, and we compare them to the predictions coming from the time-dependent perturbation theory calculations.

\subsection{Further results from perturbation theory}\label{app:res_pert_theory}

We follow the procedure laid out above in section~\ref{app:perturbation_theory}, to study which eigenmodes are excited due to a perturbation that takes the form of the external tidal force described in section~\ref{sec:tidal_force}. Throughout this appendix, the amplitude of the tidal force, as defined in eqs.~\eqref{eq:tidal_phi}--\eqref{eq:angle}, is re-scaled by a factor of $A_T = 1.4$ according to the setup of Simulation A. The perturbation theory calculation is carried out from $t=-500$ Myr to $t=500$ Myr, where $t=0$ is defined as the pericentre passage, when the perturbing force reaches its peak. We assume the amplitudes remain unchanged following $t=500$ Myr, so the halo can be evolved with fixed amplitudes based on this final timestep.

As shown in figure~\ref{fig:pert_theory_amps}, the two dominant classes of modes that are excited as a result of this perturbation are $n\neq0$, $\ell=0$ breathing modes, and $\ell=2$ quadrupole modes. The frequencies of these two dominant features can be trivially calculated using the energy difference between the ground state and the excited eigenmode interfering with it:
\begin{equation}
    \label{eq:freq}
    \omega_{ij} = \frac{E_i - E_j}{\hbar} \ .
\end{equation}
Assuming the dominant effects stem from the two excited eigenmodes with the largest amplitudes ($[n\ell m] = [1, 0, 0]$ and $[n\ell m] = [0, 2, -2]$), the periods of the breathing and the rotating modes are 220 Myr and 185 Myr, respectively.

\begin{figure}
\centering
\includegraphics[width=\textwidth]{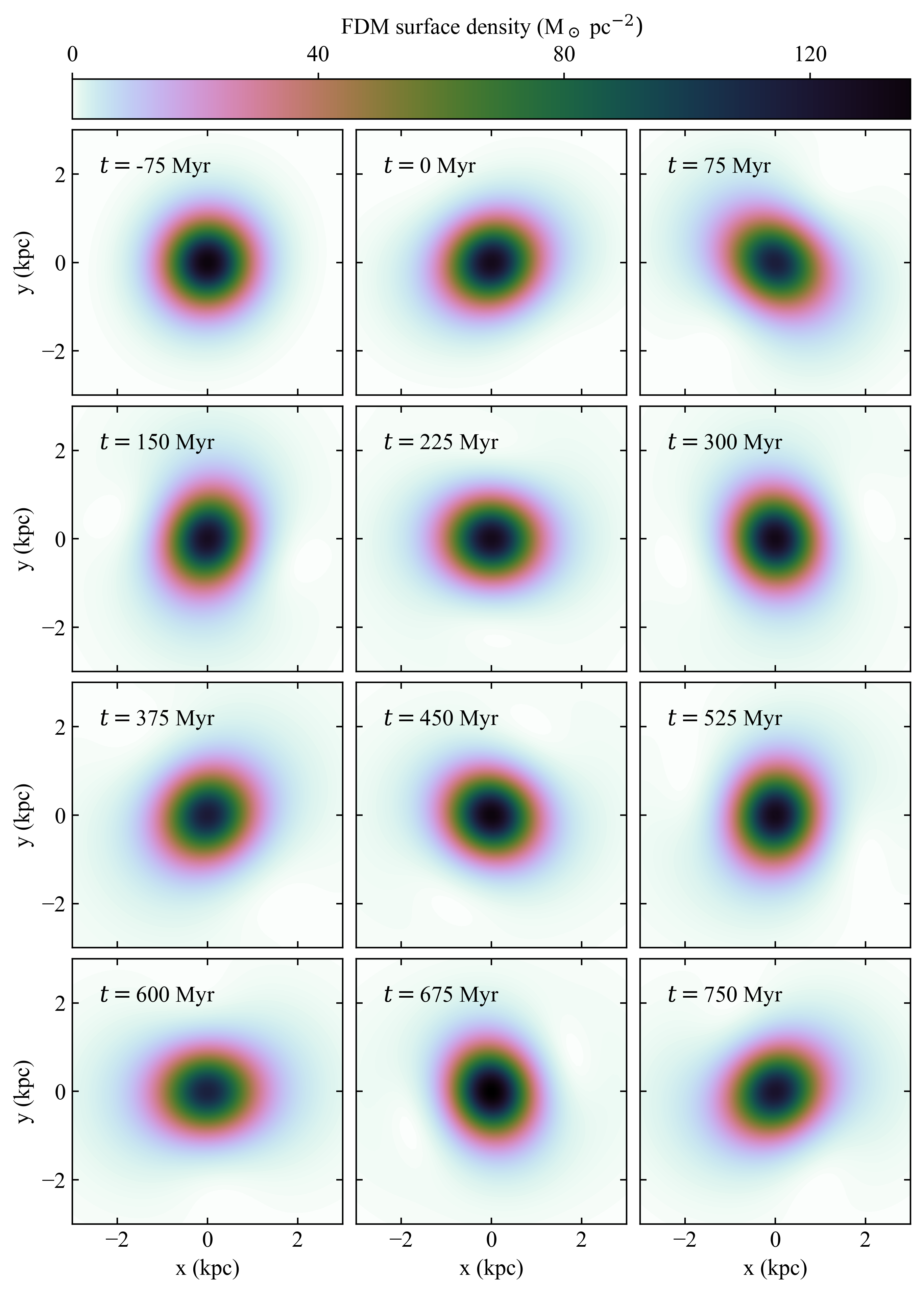}
\caption{Equivalent to figure~\ref{fig:FDM_xy_proj_A}, but for the results coming from time-dependent perturbation theory.
\label{fig:theory_dens}}
\end{figure}

\begin{figure}
\centering
\includegraphics[width=\textwidth]{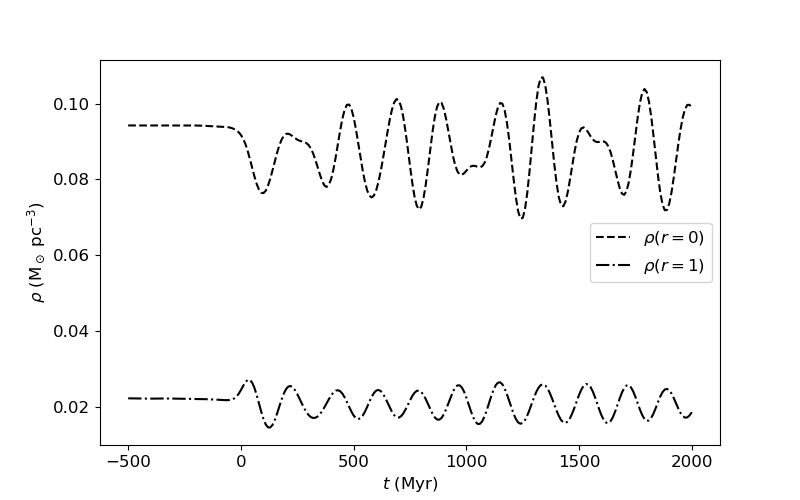}
\caption{The density as a function of time at the centre of the halo (dashed line) and at $(x,y,z) = (1,0,0)$ kpc (dot-dashed line), demonstrating the breathing and the rotating modes, respectively. The periods of the motion correspond to eq.~\eqref{eq:freq} for the interference of the ground state with the $n\ell m$ = (1,0,0) mode (top line) and the (0,2,-2) mode (bottom line). Additional modes with lower amplitudes and different frequencies complicate the signal slightly, though the dominant frequency is still evident.
\label{fig:dens_oscl}}
\end{figure}

In figure~\ref{fig:theory_dens}, we plot the projected FDM surface density in the $(x,y)$-plane given the spectrum of excited eigenmodes, for twelve different snapshots between $t=-75~\Myr$ and $t=750~\Myr$. Both the oscillation of the central density and the rotating bar-like pattern discussed above are readily apparent in this figure. Figure~\ref{fig:dens_oscl} depicts the evolution of the density at the centre of the halo and 1~kpc away from the centre at $(x,y,z) = (1,0,0)$ kpc, further demonstrating the breathing and the rotating modes described above. The periods of these modes also match the predicted periods, though the central oscillations are marked by further interference from the other $n\neq0$, $\ell=0$ states.

The main aim of the perturbation theory calculation is to build intuition for the nature of the wave-like effects that might arise in a perturbed FDM halo. Throughout this calculation, we have assumed that the overall potential that supports the wave function, and therefore the precise shape of each eigenmode, remains unchanged. This assumption is inaccurate, given the FDM wave function is itself the source of the gravitational potential, and changes to the amplitudes of the individual eigenmodes such as those described above necessarily bring about a change to the overall potential. As a result, some deviations are to be expected between the dynamical simulations and the theoretical calculation above, especially when the tidal perturbation leads to significant changes in the density profile of the halo.

\section{Supplementary simulation results}\label{app:more_results}

In this appendix, we show some supplementary results for Simulation B. We also briefly discuss the results of some alternative simulations.

\begin{figure}
\centering
\includegraphics[width=1.\textwidth]{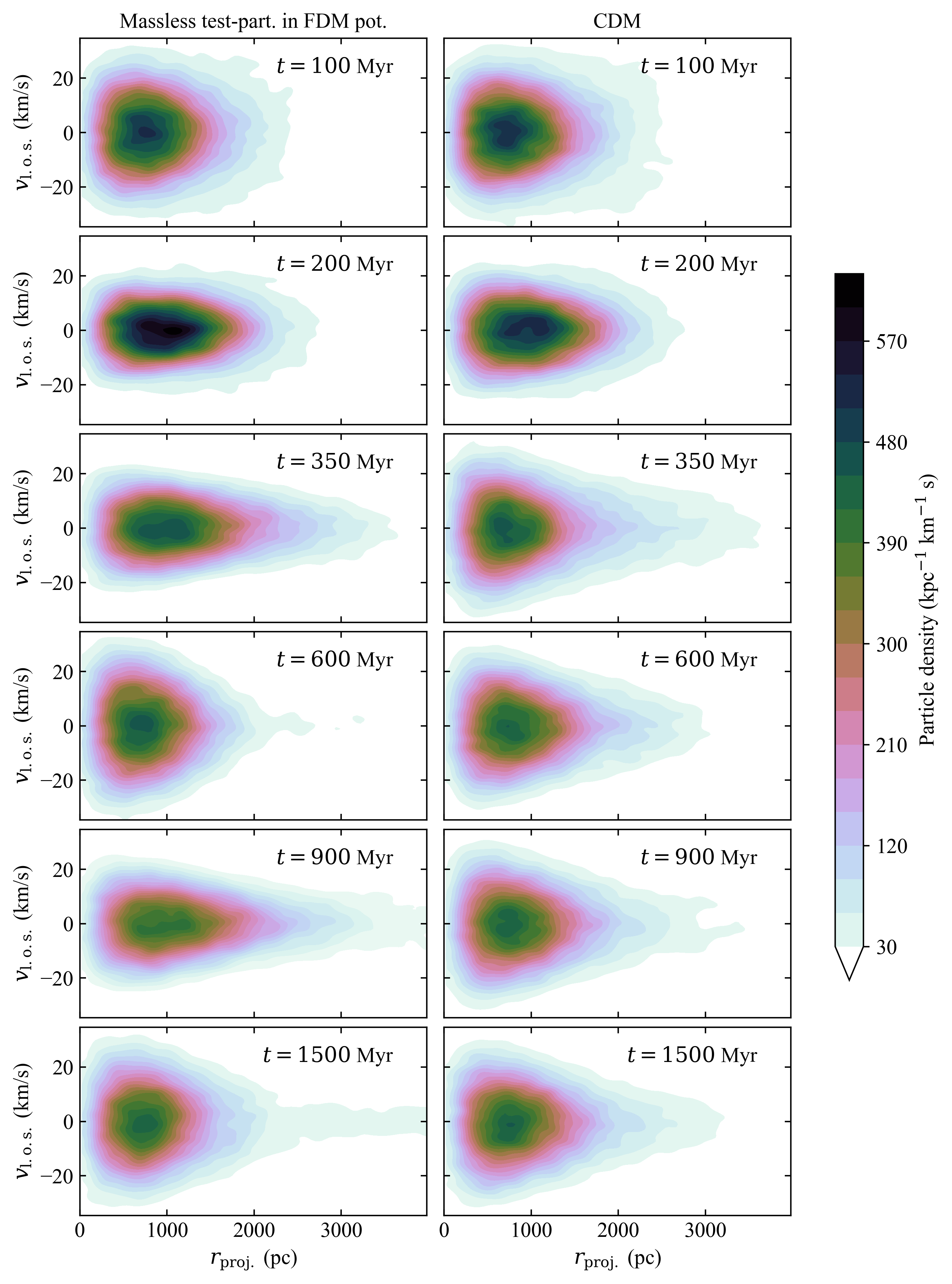}
\caption{2d histograms in the plane of projected distance ($r_\text{proj.}$) and line-of-sight velocity ($v_\text{l.o.s.}$), as observed from the centre of the host galaxy, for different snapshots in time. The two columns show a side-by-side comparison between test-particles in an FDM gravitational potential and CDM particles.}
\label{fig:projdist_vlos_2dhist_simB}
\end{figure}

Figure~\ref{fig:projdist_vlos_2dhist_simB} shows 2d-histograms of massless test-particles in an FDM potential and CDM particles of Simulation B, in the 2d-plane of projected distance ($r_\text{proj.}$) and line-of-sight velocity ($v_\text{l.o.s.}$), as defined in section~\ref{sec:coordinates}. In a scenario where proper motions are not available, these coordinates could still be available to high precision.

\begin{figure}
\centering
\includegraphics[width=1.\textwidth]{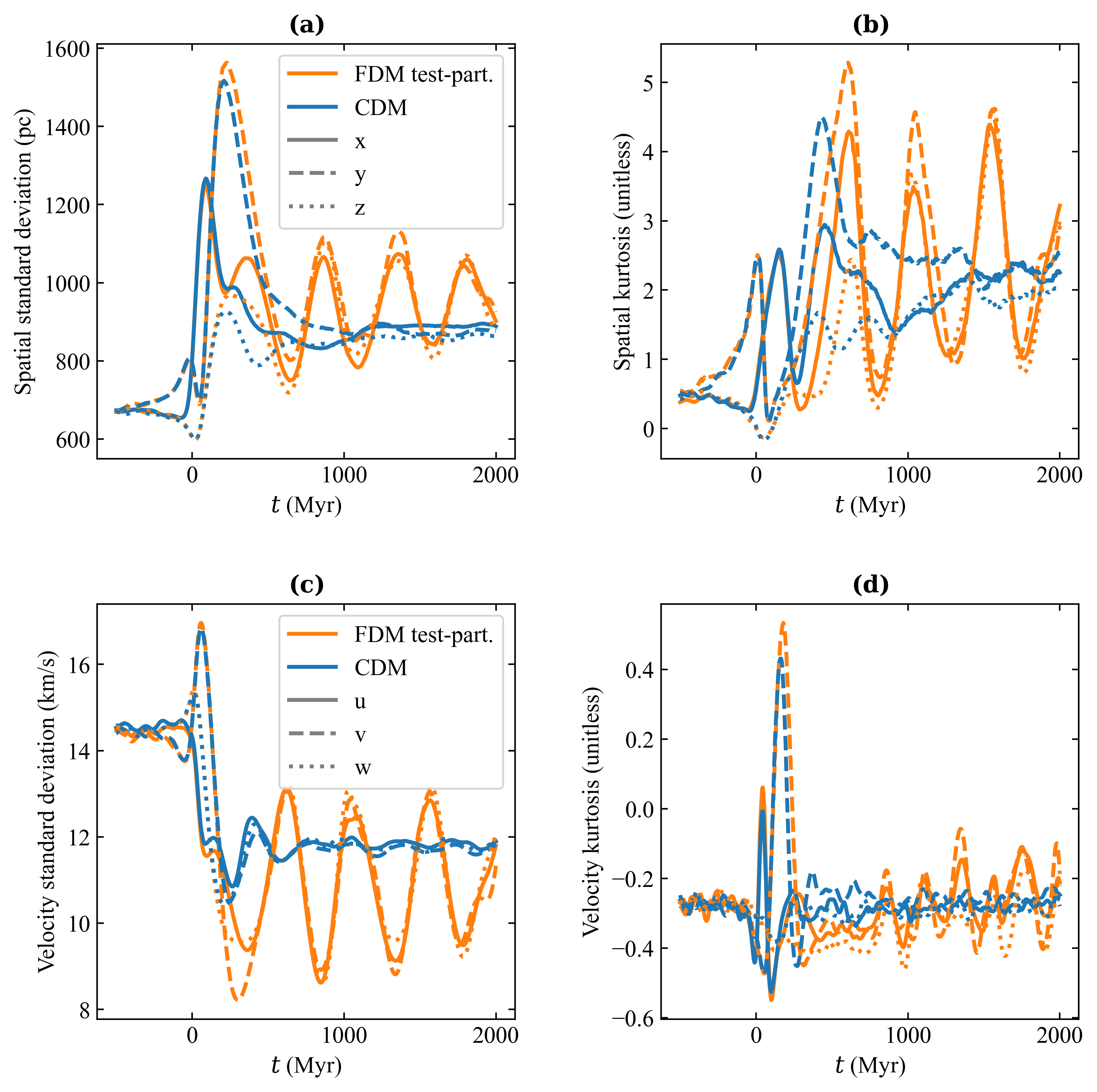}
\caption{Evolution of standard deviations (left column) and kurtoses (right column), for the spatial coordinates (top row) and velocity coordinates (bottom row), for Simulation B. The two colours represent CDM particles and massless test-particles in an FDM gravitational potential. The solid, dashed, and dotted line-styles represent the three dimensions. The kurtosis is defined according to Fischer (a standard normal distribution has a kurtosis value of zero); positive values corresponds to a more pointy distribution, with heavy tails; negative values correspond to a more boxy shape, with weak tails.}
\label{fig:std_simB}
\end{figure}

Differences in the dynamical response in Simulation B are further illustrated in figure~\ref{fig:std_simB}, showing the evolution of the phase-space distributions' standard deviations and kurtoses. Most notably, the spatial distribution of the test-particles have a kurtosis that varies quite dramatically. Similar structures were seen in both Simulation A and C. However, we choose to show Simulation B in this section, as those results were the most pronounced and clearly visible. See sections~\ref{sec:simB} and \ref{sec:discussion} for further discussion about these results.

We also ran $N$-body simulations that were initialised assuming a constant anistropy.\footnote{This is commonly parametrised by
\begin{equation}
    \beta = 1- \frac{\sigma_\theta^2+\sigma_\phi^2}{2\sigma_r^2},
\end{equation}
where $\sigma_r$ is the velocity dispersion in the radial direction and $\sigma_\theta$ and $\sigma_\phi$ are dispersions in the other two orthogonal directions. We refer to e.g. ref.~\cite{Lacroix2018} for details.}
We did not see any qualitative differences in the dynamical response of systems with this alternative initial condition.

As an additional test, we consider the case where the density distribution of massless test-particles and CDM particles are not proportional to the matter density of the FDM soliton. This is achieved after running the simulation by making cuts in the initial particle energy. By making a cut solely in total energy, meaning the sum of kinetic and potential energy, the new sub-group of particles is guaranteed to also be in a steady state initially. As described at the end of section~\ref{sec:discussion}, both of the aforementioned modifications gave rise to similar results.

\end{document}